\documentclass[usenatbib]{mn2e}
\usepackage{graphicx}
\usepackage{epstopdf}
\usepackage{amsmath}
\usepackage{amssymb}
\usepackage{times}

\usepackage{csz} % Library of personal definitions
\usepackage{bib} %Journal Abreviations, also in emulateapj
\usepackage{abrev} %astronomical abreviations
\usepackage{ifthen}
 
\begin{document}

\title[Implementing $H_2$ in Simulations]{Implementing Molecular Hydrogen in Hydrodynamic Simulations of Galaxy Formation}

\author[C. Christensen et al.]
{Charlotte Christensen,$^1$
Thomas Quinn,$^2$
Fabio Governato,$^2$
Adrienne Stilp,$^2$
\newauthor 
Sijing Shen,$^3$
James Wadsley,$^4$\\
$^1$Steward Observatory, University of Arizona, 933 North Cherry Avenue, Tucson, AZ 85721-0065\\
$^2$Department of Astronomy, University of Washington, Box 351580, Seattle WA, 98195\\
$^3$Department of Astronomy \& Astrophysics, UC Santa Cruz, 1156 High Street, Santa Cruz, CA 95064\\
$^4$Department of Physics \& Astronomy, ABB-241, McMaster Univ., 1280 Main St. W, Hamilton, ON, L8S 4M1, Canada
}

\maketitle
\begin{abstract}
Motivated by the observed connection between molecular hydrogen ($\Hmol$) and star formation, we present a method for tracking the non-equilibrium abundance and cooling processes of $\Hmol$ and $\Hmol$-based star formation in Smoothed Particle Hydrodynamic simulations.
The local abundances of $\Hmol$ are calculated by integrating over the hydrogen chemical network.
This calculation includes the gas-phase and dust grain formation of $\Hmol$, shielding of $\Hmol$, and photodissociation of $\Hmol$ by Lyman-Werner radiation from nearby stellar populations.
Because this model does not assume equilibrium abundances, it is particularly well suited for simulations that model low-metallicity environments, such as dwarf galaxies and the early Universe.
We further introduce an explicit link between star formation and local $\Hmol$ abundance.
This link limits star formation to ``star-forming regions,'' represented by areas with abundant $\Hmol$.
We use simulations of isolated disk galaxies to verify that the transition from atomic to molecular hydrogen occurs at realistic densities and surface densities.
Using these same isolated galaxies, we establish that gas particles of $10^4 \Msun$ or less are necessary to follow the molecular gas in this implementation.

With this implementation, we determine the effect of $\Hmol$ on star formation in a cosmological simulation of a dwarf galaxy.
This simulation is the first cosmological simulation with non-equilibrium $\Hmol$ abundances to be integrated to a redshift of zero or to include efficient SN feedback.
We analyze the amount and distribution of star formation in the galaxy using simulated observations of the HI gas and in various optical bands.
From these simulated observations, we find that our simulations are consistent with the observed Tully-Fisher, global Kennicutt-Schmidt, and resolved Kennicutt-Schmidt relations.
We find that the inclusion of shielding of both the atomic and molecular hydrogen and, to a lesser extent, the additional cooling from $\Hmol$ at temperatures between 200 and 5000 K increases the amount of cold gas in the galaxies.
The changes to the ISM result in an increased amount of cold, dense gas in the disk of the galaxy and the formation of a clumpier interstellar media (ISM).
The explicit link between star formation and $\Hmol$ and the clumpier ISM results in a bluer galaxy with a greater spatial distribution of star formation at a redshift of zero. 

\end{abstract}
\begin{keywords}
hydrodynamics, ISM: molecules, stars: formation, galaxies: dwarf, galaxies: evolution, galaxies: dwarf
\end{keywords}

%----------------------------------------------
\section{Introduction}  \label{introsec4}
Strong observational evidence relates star formation to molecular hydrogen, $\Hmol$.
For example, molecular clouds (MCs) are observed to be the sites of star formation both in the Milky Way \citep[e.g.][]{Blitz93} and in nearby galaxies \citep{Fukui10}.
\citet{Wong02} showed that the star formation rate (SFR) in a set of CO-bright spirals is more closely correlated to the $\Hmol$ abundance than the atomic hydrogen (HI) abundance while \citet{Fumagalli09} found that galaxies with low levels of star formation are observed to be $\Hmol$-poor.
The connection between $\Hmol$ and star formation in other galaxies has been further supported by data from THINGS \citep{Walter08}, HERACLES \citep{Leroy09}, NGS \citep{Gil07}, and SINGS \citep{Kennicutt03}.
Synthesizing these data, \citet{Leroy08} found that $\Hmol$ correlates strongly with the star formation efficiency (SFE).

% -------------- KS Law ---------------------------
The best-known example of the connection between $\Hmol$ and star formation comes from the Kennicutt-Schmidt (K-S) relation \citep{Schmidt59,Kennicutt98a}.
This law relates the surface density of star formation to the surface density of HI, $\Hmol$, or total gas and is frequently used as a benchmark for computational models of star formation.
It has been calculated both over entire galaxies  and on smaller spatial scales.
\citet{Bigiel08, Bigiel10} created a spatially resolved K-S relation using data from THINGS and related surveys by finding the SFR surface density as a function of the gas surface density on sub-kpc scales.
They found that at gas surface densities where $\Hmol$ dominates ($\Sigma_{\mathrm{gas}} \gtrsim 9 \mathrm{ M_{\sun} pc^{-2}}$) the SFE is higher than in lower surface density areas.
This change in the efficiency implies that the SFR is directly related to the surface density of $\Hmol$ and less directly to the total gas surface density.
Extending this study to greater radii in the galaxies by stacking spectra, \citet{Schruba11} verified that this relationship held out to $r_{25}$, where $\Hmol$ had been previously undetected.
Other studies of the K-S relation comparing the SFR to $\Hmol$ at various spatial scales and across a wide sample of galaxies also found that the SFR correlates most strongly with $\Hmol$  \citep{Rownd99, Murgia02, Heyer04, Gao04, Kennicutt07, Blanc09,Warren10,Bigiel10}.

% --------------- Theoretical Models -------------
Theorists have developed different analytical models relating the SFR and $\Hmol$ abundance to properties of the ISM. 
Based on the relationship between pressure and the transition from HI to $\Hmol$ presented in \citet{Elmegreen93}, \citet{Wong02} hypothesized a dependency of $\Hmol$ fraction on midplane pressure for their sample of galaxies.
Similarly, \citet{Blitz04,Blitz06a} developed a pressure-based star formation law from this relationship.
 \citet{Krumholz12} argued for a volumetric, molecular K-S relation that would account for the variation in the total-gas surface density-based K-S relation observed between spiral, dwarf, and starbursting galaxies.
 \citet{Krumholz09} used an analytic model of the equilibrium state of photodissociation fronts in MCs \citep{Krumholz08} to model the atomic to molecular transition throughout galaxies as a function of metallicity and column density. 
This model was used to develop a star formation law applicable to both primarily atomic and primarily molecular gas in \citet{Krumholz09a}.
\citet{MacLow10}, in contrast, presented a non-equilibrium, time-dependent analytic model for the $\Hmol$ fraction in a region based  primarily based on the metallicity and density.
Analytic models of star-forming regions such as these have been extensively used to study $\Hmol$ and star formation.
However, simulations of galaxies are necessary to study how $\Hmol$ evolves in concert with other processes.

% ------------------ Why sims need H2 for star formation------------------
Until recently, simulations of galaxy formation have largely neglected $\Hmol$ because of the high resolution required (see below for a description of more recent simulations that do include $\Hmol$).
Including $\Hmol$ in galaxy formation simulations, however, has the potential to affect both the SFR and the spatial distribution of star formation.
If the formation of $\Hmol$ is a fundamental step in star formation, then this omission would result in a less physical connection between the gas properties and star formation.
Linking star formation to $\Hmol$ would increase the connection between star formation and the properties of the gas that determine the fraction of $\Hmol$.
In addition to volume density, these factors may include the gas surface density, the metallicity, and the dissociating radiation field.
The distribution of both metallicity and the radiation field could affect the distribution of star formation and should potentially reflect this variation.
As will be discussed further, reproducing the correct spatial distribution of stars is important, not only for comparisons to observations, but because stars can, through supernova (SN) feedback, impact both the baryonic and dark matter (DM) mass distribution of galaxies \citep{Governato10}.
Furthermore, these properties vary across galactic type and redshift, potentially leading to varying SFRs.
For this reason, an $\Hmol$-based star formation law has been used to motivate reduced star formation efficiencies in the low-metallicity environments of dwarf galaxies and at high redshift \citep[e.g.][]{RobertsonKravtsov08,Gnedin10,Krumholz11a}.

%-------------------------- Why sims need H2 for cooling --------
A second potential effect of including $\Hmol$ is in the modeling of the cold, dense ISM.
The gas shielding necessary for $\Hmol$ formation reduces the photoheating, leading to increased amounts of cold gas.
The importance of shielding to the creation of cold, star forming-gas has been stressed in \citet{Krumholz11} and \citet{Glover11a}.
Furthermore, while $\Hmol$ does not emit strongly at the cool temperatures of MCs \citep{Glover11a}, $\Hmol$ can in low metallicity regions be an important coolant at temperatures between 5000 and 200 K through collisions \citep{Glover07, Gnedin10}.
In the early universe this cooling was responsible for the formation of Population III stars \citep{Abel97}.
At early times and in low-metallicity environments, it could also have a strong impact on the density structure of the gas and the amount of gas and star formation in DM halos \citep{Jappsen07}.

The combination of shielded gas and higher low-temperature cooling rates promotes the formation of cold, dense clouds.
Including $\Hmol$ is, therefore, important for better reproducing the environment where star formation is observed to take place.
The formation of a clumpy ISM has been shown to increase the efficiency of SN feedback at removing gas, \citep{Governato10, Brook10, Brook11a}. 
Consequently, by enabling the formation of cold, dense gas and limiting star formation to these star formation regions, $\Hmol$ has the potential to make the ISM more prone to outflows.
Outflows such as these can change both the mass and mass distribution of galaxies.
Because of this interaction between feedback and the thermodynamics of the ISM, it is necessary to include both $\Hmol$ and efficient SN feedback in simulations.

% ------------------------------- Sims that have H2 ------------------
In the past five years, several models have been developed to either study MCs or to model $\Hmol$ in simulations of galaxies.
\citet{Dobbs08, Dobbs11} and \citet{Tasker11} studied the formation, orbits, and mass distributions of MCs in models of spiral galaxies by identifying gas clumps where MCs would be.
These simulations followed the dynamics of MCs but do not include the formation or destruction of hydrogen molecules and the heating and cooling associated with them.

% ---------------------------- Prev Sims PP& R+K
Other simulations of galaxy evolution have included models for the $\Hmol$ abundance and used it to calculate the SFR.
Based on observed scaling laws for MCs \citep{Larson81}, \citet{Pelupessy06} presented a time-dependent sub-grid method for simulating MCs in Smoothed Particle Hydrodynamic (SPH) simulations.
This method has been used to reproduce the effect of metallicity on the $\Hmol$ abundance in dwarf galaxies \citep{Pelupessy06}, the relationship between $\Hmol$ abundance and mid-plane gas pressure, and the conversion factor between $\mathrm{CO}$ and $\Hmol$ \citep{Pelupessy09}.
Because $\Hmol$ was linked to star formation, this model has also been used to study deviations from the K-S relation in low-metallicity, high-redshift galaxies \citep{Papadopoulos10}.
\citet{RobertsonKravtsov08} also tracked the local $\Hmol$ abundance in their simulations of galaxies.
For their model, they used the equilibrium abundance of $\Hmol$ for gas particles, based on the metallicity of the gas and the amount of dissociating radiation. 
With these abundances, they were able to examine the importance of $\Hmol$ to star formation in their SPH simulations of isolated galaxies and reproduce deviations from the global K-S relation in low metallicity galaxies based on reduced $\Hmol$ fractions.
\citet{Kuhlen11} applied the equilibrium models for $\Hmol$ abundance from \citet{Krumholz09} to models of dwarf galaxies at high z and found that in the absence of efficient SN feedback, $\Hmol$-based star formation reduced the stellar mass in dwarf galaxies.
Each of these models, however, rely either on assuming the global properties of MCs or assuming that the gas is in chemical equilibrium.

% --------------------------- Prev Sim G+K
In the only fully self-consistent implementation so far, \citet{Gnedin09} implemented a model for the local non-equilibrium $\Hmol$ abundances in their Adaptive Mesh Refinement (AMR) simulations.
In this model, they used local rates of $\Hmol$ formation and destruction to integrate over the chemical network.
Using this method, they created cosmological simulations of galaxies to determine deviations from the K-S relation in low metallicity galaxies \citep{Gnedin10} and with different dust-to-gas ratios and far-UV fluxes \citep{Gnedin11} and to study sources of scatter in the K-S relation \citep{Feldmann11}.
This model has since been used to verify equilibrium analytic models for the $\Hmol$ abundance for gas with metallicity $\geq 10^{-2}$ Z$_{\sun}$ \citep{Krumholz10}.

%----------------------------- Reason?
In this paper, we have adapted the method presented in \citet{Gnedin09} for SPH simulations with efficient SN feedback.
We present a method for integrating the local non-equilibrium abundance of $\Hmol$ in SPH simulations based on the local formation and destruction rates.
These rates include the formation of $\Hmol$ on dust grains and photodissociation of $\Hmol$ by Lyman Werner (LW) radiation from nearby stellar populations.
The calculation of the local LW flux is based on the radiation from nearby stellar particles, in which the gravity tree structure is used to establish proximity.
Accordingly, we also model the shielding of $\Hmol$ from the LW radiation by both itself and dust for each gas particle.
We also include the dust shielding of atomic hydrogen from the cosmological UV background.
Molecular hydrogen-based star formation and a blastwave model for SN feedback are also included in the simulations, in order to produce both star formation and SN feedback in regions with high $\Hmol$ abundances.
This model is ideally suited for simulations of galaxy formation because of the speed and ability to simulate large spatial ranges inherent in SPH.
By integrating over the chemical network, no assumptions are made about the final structure of molecular clouds or the relationship between $\Hmol$ and global galactic properties.
It is, therefore, especially useful for studying extreme environments, such as the early Universe and low-metallicity dwarf galaxies.

%------------------ What we do ----------------
Despite the work done on $\Hmol$ in evolving galaxies, few of the previous simulations with $\Hmol$ have been cosmological and none have been integrated to a redshift of zero.
Furthermore, none of them have included both $\Hmol$ and efficient feedback.
Cosmological simulations are necessary to determine the effect of $\Hmol$ on star formation  and disk structure in realistic models of galaxy formation that include gas accretion through both cold-flows and mergers.
Following these simulations to a redshift of zero is necessary to analyze the effect of $\Hmol$ on the stellar and gaseous content of galaxies over the history of the Universe and to compare them to observations of nearby galaxies.
The inclusion of efficient feedback is important for correctly modeling star formation and gas loss and for examining the connection between the formation of cold, dense gas and outflows.
For these reasons, we present a cosmological simulation of a dwarf galaxy with non-equilibrium abundances of $\Hmol$ and efficient feedback integrated to z=0.

%------------------ Why Dwarfs -----------------
We focus this work on field dwarf galaxies, both because of their manageable computational scale and scientific interest.
Dwarf galaxies are well suited for simulations because they require fewer particles to achieve high mass resolution.
Their low virial masses and metallicities also make them sensitive laboratories for studying gas dynamics and star formation.
Having a small virial mass means that gas heated through SN feedback can more easily escape the halo.
Their low metallicities reduce the cooling rates of the gas, making $\Hmol$ a relatively more important coolant.
Their relative lack of dense gas also makes these galaxies an extreme environment for star formation.
Their SFE may be even further reduced by the difficulty of $\Hmol$ formation in such a low metallicity and low surface density environment.
With their low metallicities, the formation of $\Hmol$ and, consequentially, star formation will occur at higher densities than in spiral galaxies.

%-------------- outline ---------------------
This method may be modified for other SPH simulations and we anticipate that our results will be of interest to those concerned with modeling the connection between ISM and star formation and to those interested in star formation in dwarf galaxies.
In $\S$\ref{sec:methods}, we describe our algorithms for calculating $\Hmol$ abundance and simulating star formation and test our model on isolated disk galaxies.
In $\S$3 we present the results comparing the effect of different models of the ISM and star formation recipes on the amount and distribution of star formation.
Finally, in $\S$4, we discuss how the changes in the star formation were related to changes in the thermodynamics of the gas and discuss the connection to other works.

%-%%%%%%%%%%%%%%%%%%%%%%%%%%%%%%%%%%%%%%%%%%%%%%%%%
\section{Methods: Molecular Hydrogen Implementation}\label{sec:methods}
We implement a method for tracking the local, non-equilibrium abundance of $\Hmol$ in our SPH simulations based on the local formation and destruction rates.
The abundance of $\Hmol$ is dominated by the balance between dust grain formation ($\S$~\ref{sec:dust}) and photodissociation rates.
We estimate the photodissociation rate based on the LW flux from surrounding star particles ($\S$~\ref{sec:photo}).
The gas is in turn shielded from LW radiation by itself and dust ($\S$~\ref{sec:shielding}).
Gas-phase reactions ($\S$~\ref{sec:gasphase}) also impact the abundance of $\Hmol$ in low-metallicity environments and provide an additional source of cooling.
We model each of these processes and use a semi-implicit solver to calculate the $\Hmol$ abundance in each gas particle at each time-step.
This implementation for calculating $\Hmol$ abundances is based on the approach of  \citet{Gnedin09} but with a simplified estimate for LW radiation and applied to SPH, rather than AMR, simulations.
These simulations also include a more efficient SN feedback model.
It is the first self-consistent non-equilibrium $\Hmol$ implementation to be used in conjunction with efficient SN feedback and in cosmological simulations integrated to $z = 0$.

%------------------------- How we will test it ---------------
We tested our implementation on isolated models of a Milky Way-mass galaxy ($\S$~\ref{sec:test} and $\S$~\ref{sec:res}) before completing the cosmological dwarf galaxies.
We, therefore, assess our $\Hmol$ implementation over a range of masses in this paper.
The initial conditions for the isolated simulations are described in Appendix A.
In particular, we ensured that the transition from atomic to molecular hydrogen occurred at the correct densities and surface densities at our resolution and displayed a realistic metallicity dependency.
This transition is an important constraint because the presence of $\Hmol$ is observed to be strongly correlated with the surface density of gas.
Furthermore, many theoretical models for the abundance of $\Hmol$ primarily depend upon the extent that the gas is shielded, a quantity that is directly related to surface density.
We used these same simulations to determine the necessary resolution to model $\Hmol$ and the conditions in which it forms.

\subsection{General Description of Code}
\label{sec:general}
%-------------------------- general description of Gasoline
We implement $\Hmol$ physics in the galaxy formation code, GASOLINE.
GASOLINE is a parallel tree+SPH code \citep{Wadsley04} and an extension of the \emph{N}-body gravity code PKDGRAV \citep{Stadel01}.
GASOLINE has been used extensively for studying galaxy formation and evolution and has been successful in reproducing the Tully-Fisher Relation \citep{Governato09},  the rotation curves of dwarf galaxies \citep{Governato10}, the internal structure of disk galaxies \citep{Governato07, Roskar08, Stinson09, Brooks09}, the column density distributions of damped Lyman-$\alpha$ systems \citep{Pontzen08,Pontzen10}, and the observed stellar mass-metallicity relationship \citep{Brooks07}.

%--------------------- What chemistry gasoline includes -------------------
In addition to gravity and gas dynamics, GASOLINE models physical processes such as ion abundances, background UV, gas cooling rates,  star formation, and feedback from SNe.
We use a semi-implicit integrator to calculate the non-equilibrium ion abundances and gas cooling from collisional ionization rates \citep{Abel97}, radiative recombination \citep{Black81, VernerANDFerland96}, photoionization, bremsstrahlung, and cooling from H, He \citep{Cen92} and metal lines \citep{Shen10}.
For the purposes of calculating the rates of radiative recombination and photoionization, our simulations include a redshift-dependent UV-background radiation from quasars and stars in external galaxies \citep{Haardt96}.
Cooling rates from metal lines are tabulated based on the gas temperature, density, and metallicity and the UV background.
These rates were computed using CLOUDY \citep[version 07.02;][]{Ferland98} with the assumptions that the gas is in ionization equilibrium and optically thin to UV radiation.
In this work, we extended these calculations to include the presence of $\Hmol$, as described in \S\ref{sec:H2imple}.

%---------------------- SF and SN feedback in gasoline
Star formation in GASOLINE is determined on a probabilistic basis according to the free-fall time of potentially star-spawning gas particles \citep{Stinson06}.
We introduce a modification to our star formation recipe, further described in \S\ref{sec:H2SF}, in which the abundance of $\Hmol$ is incorporated into the star formation probability.
In order to calculate the effects of feedback from Type Ia and Type II SNe, we use the blastwave method outlined in \citet{Stinson06}.
In this method, energy from SNe is distributed among surrounding gas particles and gas cooling is disabled for a length of time and distance determined by the blastwave model.
This feedback method is very efficient at producing gas outflows.
Supernovae are also responsible for enriching the surrounding gas particles and metals are distributed throughout the simulation through metal diffusion \citep{Shen10}.

\subsection{Implementation of  Molecular Hydrogen}
\label{sec:H2imple}
%--------Introducing H2 implementation --------------------------------------------
%-------- Chemical Evolution ------------------------------
The abundance of $\Hmol$ is based on the rates of dust grain $\Hmol$ formation, photodissociation with shielding, gas-phase $\Hmol$ formation, and collisional dissociation.
The abundance of HI is now a function of the rate of $\Hmol$ formation, in addition to the rates of photoionization, recombination, and collisional dissociation.
Using these rates, the equations of balance for HI and $\Hmol$ may be written as 

\begin{align}
& \dot{X}_{\mathrm{HI}} = R(T)n_e X_{\mathrm{HII}} - S_d X_{\mathrm{HI}}n_b \Gamma_{\mathrm{HI}} - C_{\mathrm{HI}}n_{\mathrm{HI}}n_e - 2 \dot{X}_{\Hmol}\\
& \dot{X}_\Hmol = R_d n_b X_{\mathrm{HI}}(X_{\mathrm{HI}} + 2 X_{\Hmol}) - S_d S_\mathrm{H_2} X_\Hmol n_b \Gamma_\Hmol^{\mathrm{LW}}  +  \dot{X}^{gp}_\Hmol
\end{align}
where $X_i$ is the number fraction of baryons in species $i$
\begin{equation}
X_i = n_i/n_b,
\end{equation}
$R(T)$ is the recombination rate, $n_e$ is the number density of electrons, $S_d$ is shielding from dust, $n_b$ is the number density of baryons, $\Gamma_{\mathrm{HI}}$ is the hydrogen photo-ionization rate, $C_{\mathrm{HI}}$ is the HI collisional ionization rate, $R_d$ is the formation rate coefficient of $\Hmol$ on dust, $S_\Hmol$ is self-shielding of $\Hmol$, $\Gamma_\Hmol^{\mathrm{LW}}$ is the $\Hmol$ photodissociation rate, and $X^{gp}_\Hmol$ is the gas-phase dissociation and formation rate of $\Hmol$.
The shielding factors, $S_d$ and $S_\Hmol$, represent the fraction of ionizing or dissociation flux that reaches the gas and range from zero to one.
The gas-phase dissociation and formation rate includes collisional dissociation of $\Hmol$ and the gas-phase formation of $\Hmol$ via $\mathrm{H}^-$.
We describe the dust-based formation, the rate of photodissociation and the shielding, and the gas-phase dissociation and formation in the following sections.

\subsubsection{Formation on Dust Grains}
\label{sec:dust}
The formation of $\Hmol$ on dust grains dominates over gas-phase formation once a small reservoir of metals is available.
As in \citet{Gnedin09}, we use an observationally motivated equation for the coefficient for rate of $\Hmol$ formation on dust, $R_d$, from \citet{Wolfire08}.
In this formalization, the rate of $\Hmol$ formation on dust depends on both the amount of dust and the gas density structure.
In order to calculate the amount of dust, we assume a dust-to-gas ratio proportional to the gas metallicity, Z.

The sub-grid gas density structure affects the rate of $\Hmol$ formation through the increased rate of formation in unresolved high-density regions. 
Therefore, any model that follows the non-equilibrium abundances of $\Hmol$ without resolving the turbulent structure within gas clouds must somehow parametrize the structure.
We address the sub-grid nature of $\Hmol$ formation in the same way as \citet{Gnedin09}, by assuming that the gas has a log-normal density substructure.
On scales smaller than gas particles, the increased formation rate caused by the substructure can be characterized by a clumping factor, $C_{\rho}$, equal to $\langle\rho^2\rangle/\langle\rho\rangle$.
For a log-normal density distribution $C_{\rho} = e^{\sigma^2_{ln_{\rho}}}$, in which $\sigma_{\rho}$ is the width of the distribution. 
Within our simulations $C_{\rho}$ is an adjustable parameter with a best-fit value of 10.0, as in \citet{Gnedin09}.

The accuracy of estimating the formation enhancement with this clumping factor has been confirmed by \citet{Milosavljevic11} for $\Hmol$ fractions up to 50\%.
However, \citet{Milosavljevic11} also cautioned it may over-predict the rate of formation at higher densities and $\Hmol$ fractions.
Therefore, further development to the equation for $\Hmol$ formation at high densities and molecular fractions may be warranted.
However, we choose to focus our analysis on situations in which fully molecular gas is rare (dwarf galaxies) and where the transition from mostly atomic to predominantly molecular gas is more important than the exact $\Hmol$ abundance at very high densities.
Including both the effect of dust and sub-grid clumping, our $\Hmol$ formation rate is defined as
\begin{equation}
R_d = 3.5 \times 10^{-17} Z/Z_{\sun} C_{\rho} \mathrm{~cm^3 s^{-1}}
\end{equation}
 
\subsubsection{Photodissociation by Lyman Werner Radiation}
\label{sec:photo}
Dissociation of $\Hmol$ is caused primarily by LW radiation ($\Gamma_\Hmol^{LW}$), which can either be produced by young stars in the galaxy or can come, to a much smaller degree, from extra-galactic sources.
We assume the LW radiation from extra-galactic sources takes the form of a cosmological background varying with redshift and use the function from \citet{Haardt96}.
At $z=0$, this radiation is approximately $5000$ times less than the average flux of LW radiation in the disks of galaxies.  
For $z \geq 2$, the background radiation approaches 50 times less than the average flux. 
We include cosmic LW radiation for consistency but it has a negligible impact on the star formation, as also observed in \citet{Gnedin10a}.

Stellar light in the galaxy is the primary source of LW radiation.
This radiation depends on the distribution of young stars and can only be correctly determined using full radiative transfer.
Unfortunately, full radiative transfer is currently computationally prohibitive in cosmological simulations that are evolved for a Hubble time.
Even those cosmological simulations that do include radiative transfer must make assumptions, such as using an optically thin approximation to find the Eddington tensor \citep{Gnedin01, Petkova09}.

For the sake of speed and computational ease, we make a simple approximation of the amount of stellar LW flux a given gas particle experiences based on the average flux from nearby stars.
When calculating the flux, we assume a completely optically thin disk.
However, as will be shown in $\S$~\ref{sec:shielding}, we do include shielding from both dust and $\Hmol$ molecules when calculating the photodissociation rate of $\Hmol$ for a gas particle.
This method for calculating the LW flux relies on LW dissociation being primarily important over short distances and the absorption of LW photons occurring mostly in MCs.
Fortunately, the efficient shielding of  $\Hmol$ at high densities causes the fraction of $\Hmol$ to be relatively insensitive to the amount of dissociation radiation incident on MCs \citep{Gnedin09,Krumholz09,MacLow10}.
As described below, this method is sufficient to follow the large-scale spatial and temporal variations of the LW radiation and to reproduce the surface density at which hydrogen becomes molecular to within a factor of two.

The details of our method for calculating the LW flux are as follows.
The average flux impacting a gas particle from nearby stars is calculated using the tree built for the gravity calculation.
To do this, we apply the flux from star particles within the same cell or, if necessary, the same parent cell in the following manner.
Within each cell of the tree, we assume that the amount of LW flux coincident on a given gas particle is a function of the amount of LW radiation emitted by star particles within the same cell. 
Each star particle emits LW radiation ($L$) according to its age ($t$) and mass ($m$), based on the Starburst99 \citep{Leitherer99} calculation for a simple stellar population with a \citet{Kroupa93} IMF.
We then define one average LW source for each cell with luminosity, $L_s$, equal to the total luminosity of stars within the cell and located at the luminosity-weighted average location, $\mathbf{x_s}$, of the star particles.
Based on the luminosities of individual star particles and their locations ($\mathbf{x}_i$), we define $L_s$ and $\mathbf{x_s}$ as

\begin{align}
& L_s = \sum_{i=0}^{N} L(t_i,m_i)\\
& \mathbf{x_s} = \frac{\sum_{i=0}^N L(t_i,m_i) \mathbf{x_i} }{L_s}
\end{align}

Gas particles within the cell experience a LW flux as if all radiation were coming from the average LW source.
For a gas particle with a distance greater than one softening-length away from the average source, we set the flux equal to what it would be from a point source at that distance and with that luminosity.
For a gas particle with a distance less than the gravitational softening, we set the flux equal to the flux at the softening radii.
The flux, $F$, incident on a gas particle at position $\mathbf{x_g}$ is, therefore,  

\begin{equation}
F =\left \{ 
\begin{array}{rl}
\frac{L_s}{4 \pi |\mathbf{x_s} - \mathbf{x_g}|^2} &  \mbox{if $|\mathbf{x_s - \mathbf{x_g}|} \geq h$} \\
\frac{L_s}{4 \pi h^2}                                                 & \mbox{otherwise} \\ 
\end{array} 
\right .
\end{equation}
where $h$ is the softening.
This adjustment prevents artificially high levels of radiation at distances too small to be resolved.

One complication to this method arises when there is a bright LW source nearby but not within the cell.
If the radiation from this source were ignored, the amount of LW radiation on the gas particles in the cell would be underestimated.
In order to account for cases in which bright sources in adjacent cells should impact the amount of flux, we make the following adjustment.
We use the average LW source of the parent cell if doing so results in the gas generally experiencing higher amounts of LW radiation.

In order to estimate whether the radiation would be generally higher using the parent cell average source, we calculate the location and luminosity of the average sources for both the parent and child cell.
We then determine what the flux from the parent average source would be at the center of mass of the gas in the child cell.
We compare this to the flux from the child average source at a distance equal to the first moment of the gas mass in the child cell.
If the flux is higher in the first case, we calculate the amount of radiation experienced by each gas particle in the child cell based on the average LW source of the parent cell.
Otherwise, we only use the data for the child cell and proceed as in equation 7.

\begin{figure}
\centering
\includegraphics[width = .5\textwidth]{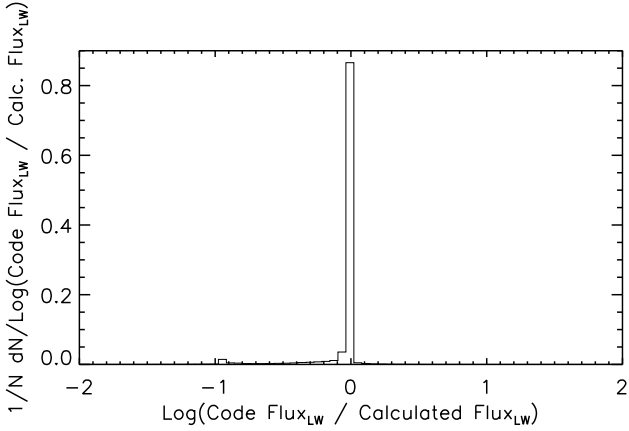}  
\caption[Accuracy of LW Flux Calculation]
{Comparison between our approximation of the LW flux experienced by each gas particle in simulation MWmr (Code $\mathrm{Flux_{LW}}$) and the amount of flux each gas particle would experience in the optically thin limit  (Calculated $\mathrm{Flux_{LW}}$).
The histogram is normalized by the total number of particles.} 
\label{fig:lw}
\end{figure}

We illustrate the accuracy of our method using a simulation of an isolated Milky Way-mass galaxy  (MWmr from Appendix A).
Figure~\ref{fig:lw} compares the LW flux at each gas particle calculated within the code to the theoretical LW flux, assuming an optically thin ISM.
\footnote{To calculate the theoretical amount of LW flux for comparison, we first calculated the amount of LW each stellar particle emitted based on the Starburst99 \citep{Leitherer99} value for simple stellar population with the same initial mass function (IMF), mass, and age. Then, for each gas particle, we summed the amount of flux from each stellar particle at the gas particle's location, assuming there was no intervening absorption.}
In our simulations, this method produced fluxes within a factor of ten of the optically thin result.
This factor of ten equates to a change of about 50\% to the density and surface density at which the transition from HI to $\Hmol$ takes place.
Despite the limitations of our method, it is able to track general trends in the LW radiation caused by different rates of star formation.

\subsubsection{Shielding from Radiation}
\label{sec:shielding}
Shielding from LW radiation is the primary reason molecular gas can form and remain in high-density regions.
We used a phenomenological model to approximate the computationally expensive three-dimensional treatment of shielding based on the surface density of each gas particle, following the work of \citet{Draine96}, \citet{Glover07} and \citet{Gnedin09}.
To fully reproduce the shielding would require a three-dimensional treatment of radiative transfer and is beyond the computational capabilities of our cosmological simulations.
However, this simplified model is able to reproduce the observed efficiency of shielding from LW radiation as a function of surface density and metallicity.

In this model, the amount the gas is shielded from LW radiation is a function of the gas surface density and metallicity (used to calculate the dust shielding) and $\Hmol$ surface density (used to calculate the self-shielding). 
We also apply dust shielding to HI when calculating the rates of photoionization and heating from the UV background radiation.
The functional forms for the dust shielding, $S_d$, and self-shielding, $S_{\mathrm{H_2}}$, are based on observations and the parameters were tuned to fit observational data.
%\begin{equation}
\begin{align}
& S_d = e^{-\sigma_{d_{eff}} Z/Z_{\sun} (N_{\mathrm{HI}} + 2N_{\mathrm{H_2} })} \\
& S_{\mathrm{H_2} } = \frac{1 - \omega_{\mathrm{H_2} }}{(1 + x)^2} + \frac{\omega_{\mathrm{H_2} }}{(1 + x)^{(1/2)}}e^{-0.00085(1 + x)^{1/2}}
\end{align}
%\end{equation}
In the above equations, $N_{\mathrm{HI}}$ is the column density of HI, $N_{\Hmol}$ is the column density of $\Hmol$, and $x = N_{\mathrm{H_2} }/(5 \times 10^{14})~\mathrm{cm^2}$. 
Both $\sigma_{d_{eff}}$ and $\omega_{\mathrm{H_2} }$ are adjustable parameters tuned to be $4 \times 10^{-21} \mathrm{cm^2}$ and $0.2$, respectively.

The HI and $\Hmol$ column densities are calculated based on the volumetric number density of a gas particle, $n_b$, and an assumed column length of the particle.
We set the column length for both dust and self-shielding to be equal to the smoothing length, $h$, of the gas particle, resulting in $N_{i} = X_{i}~h~n_b$, where $i$ may be either $\mathrm{HI}$ or $\Hmol$.
Our use of $h$ for the column length was largely decided on for empirical reasons (see $\S$~\ref{sec:test} for a test of our $\Hmol$ calculation, including this assumption of column length).
However, it also has the advantages of simplicity and being directly related to the volume over which the gas particle mass is spread.
We also investigated the use of more physically motivated equations of the column length, including the use of a turbulent scale length based on the local shear, as presented in \cite{Pavlovski02}.
We found, however, that in SPH simulations, basing the column length of a gas particle on the shear across a particle resulted in unphysically small column lengths at the densities where $\Hmol$ becomes abundant.

\subsubsection{Gas-Phase Dissociation and Formation}
\label{sec:gasphase}
%---------------- gas-phase stuff ------------
Gas-phase reactions, including both collisional formation and dissociation, can be important for establishing the abundance of $\Hmol$ in low-metallicity environments and can provide an additional source of cooling for the gas.
Gas-phase formation of $\Hmol$ occurs via $\mathrm{H_2^+} + \mathrm{H} \rightarrow \mathrm{H_2} + \mathrm{H^+}$ and $\mathrm{H^-} + \mathrm{H} \rightarrow \mathrm{H_2} + \mathrm{e^-}$.
Following the work of \citet{Abel97}, we assume that the abundance of $\Hmol$ is largely independent from the abundance of $\mathrm{H^+}$ and use equation 26 from their minimal model for $\Hmol$ formation via $\mathrm{H^-}$.
In addition to photodissociation, $\Hmol$ may be dissociated through collisions with $\Hmol$, HI, HII, and $e^-$.  
We use the rates for these collisions given in \citet{Lepp83}, \citet{Dove89}, \citet{Abel97}, and \citet{Donahue91}, respectively.
These processes together result in 
\begin{equation}
\dot{X}^{gp}_{\mathrm{H_2}} = k_1 n_{\mathrm{H^-}}n_{\mathrm{H}} - n_{\mathrm{H_2}}(k_2n_{\mathrm{H_2}} + k_3n_{\mathrm{H}} + k_4n_{\mathrm{HII}} + k_5n_e)
\end{equation}
where $n_{\mathrm{H^-}}$ is as defined in equation 27 of \citet{Abel97}, $k_1$ is the $\Hmol$ formation rate via $n_{\mathrm{H^-}}$, and $k_2$, $k_3$, $k_4$, and $k_5$  are the collisional dissociation rates by $\Hmol$, $\mathrm{H}$,  $\mathrm{HII}$, and electrons, respectively.
The values for $k_3$, $k_4$, and $k_5$ are summarized in \citet{Abel97} as reactions 11, 12, and 13, respectively. 
For all gas-phase calculations, we assume a constant ratio of ortho-hydrogen (nuclear spin number $I = 0$) to para-hydrogen ($I = 1$) equal to 3:1.
This assumption is justified in \citet{Glover08}.

\subsubsection{Heating and Cooling}
\label{sec:heatcool}
With the implementation of $\Hmol$, we added cooling of $\Hmol$ due to collisional dissociation and collisionally excited line emission.
Collisional reactions are particularly important for the cooling of $\Hmol$  when 200 K $\textless$ T $\textless$  5000 K and $\rho \gtrsim$ 10 amu/cc) \citep{Gnedin10}.

Dissociative cooling due to collisions with $\Hmol$, $\mathrm{H}$,  $\mathrm{HII}$, and electrons is calculated using the rates and amount of emission presented in \citet{Abel97}.
The collisional induced line-excitation cooling rates are determined for interactions between $\Hmol$ and $\Hmol$, $\mathrm{H}$,  $\mathrm{HII}$, HeI, and electrons.
For these interactions, we use the cooling rates presented in \citet{Glover07} for the low-density case and summarized in Table 8.
The line-excitation cooling rates from \citet{Glover07} use values calculated in \citet{Wrathmall07} from the \citet{Mielke02} potential surface and are substantially greater than the previously determined rates of \citet{Galli98}.
Deuterium cooling is ignored as it is primarily important at densities too high to be resolved in our simulations.
In addition to photoionization heating for atomic and ionized hydrogen and helium, we include heating due to photodissociation of $\Hmol$. 

\subsection{Testing Molecular Hydrogen Abundances}
\label{sec:test}
\begin{figure}
\centering
\includegraphics[width = 0.5\textwidth]{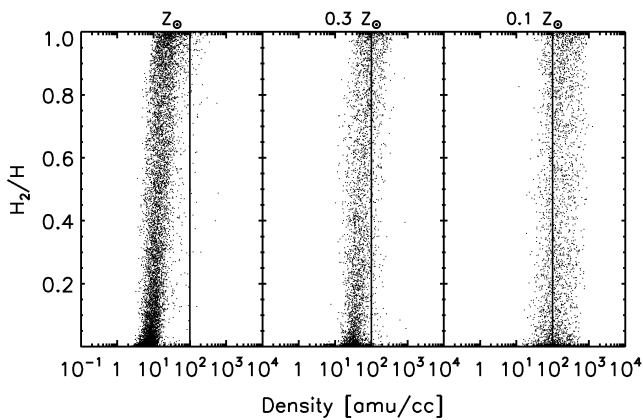}
\caption[Dependency of $\Hmol$ abundance as a function of density on metallicity]
{The transition from atomic to molecular hydrogen for three isolated, Milky Way-like simulations with different metallicities.  
The points represent individual SPH gas particles.
Vertical lines at 100 amu/cc (an average MC density and the minimum density for star formation in our non-$\Hmol$ simulations) are included to guide the eye.
As the metallicity decreases, the combination of longer formation times and less shielding causes $\Hmol$ to form at increasingly high densities. }
\label{fig:metabund} 
\end{figure}

In order to verify the accuracy of our $\Hmol$ abundance methodology, in particular the shielding calculation, we compared the volume densities and surface densities at which gas transitioned between atomic and molecular in our isolated simulations of different metallicities.
We found that the transition occurred at densities between $\rho = 10~\mathrm{amu/cc}$ and $100~\mathrm{amu/cc}$, depending on the metallicity, as predicted (Figure~\ref{fig:metabund}).
These values are consistent with average densities of MCs.

\begin{figure}
\centering
\includegraphics[width = 0.5\textwidth]{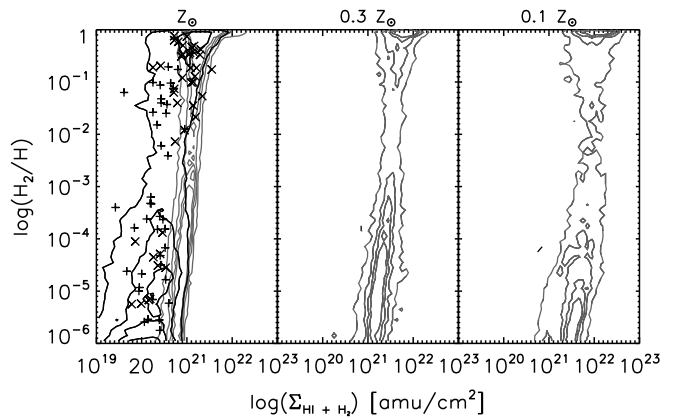}
\caption[Dependency of $\Hmol$ abundance as a function of surface density on metallicity]
{Transition from atomic to molecular hydrogen as a function of surface density for three isolated, Milky Way-like simulations with different metallicities.  All the contours represent data from the simulations. The crosses are data from  \citet{Gillmon06} and the X's are data from \citet{Wolfire08}.  The grey contours were calculated assuming surface densities of $N_{i} = X_{i}~h~\rho$, as used in the code.  
They represent the number of points (individual gas particles) enclosed and are linearly spaced.  
The black contours in the left plot were calculated to mimic observations of MCs.  
For these contours, the surface densities were calculated for each cell in a $2.5\pc \times 2.5\pc \times 10$ km/s velocity cube, in which 10 km/s was chosen to be an average velocity dispersion of a MC.
These contours represent the distribution of the individual cells in surface density and $\Hmol$-fraction space and are linearly spaced at the same levels as the grey contours.
The transition between atomic and molecular hydrogen in the solar metallicity simulation fits with the observational data.  As metallicity decreases, the transition moves to higher column densities as both dust shielding and the rate of $\Hmol$ formation on dust decrease.}
\label{fig:metabundSD}
\end{figure}

The surface density at which the transition occurs is shown in Figure~\ref{fig:metabundSD} and compared to observations from  \citet{Gillmon06} and  \citet{Wolfire08}.
When determining the dependence of the $\Hmol$ abundance on surface density, we show the surface density used in the code calculation ($N_{i} = X_{i}~h~n_b$) (grey contours) and, in the case of MWmr, the surface density as it would be determined observationally by the method used in \citet{Gillmon06} and  \citet{Wolfire08} (black contours).
In these studies, the $\Hmol$ surface density was determined from the amount of absorption in the quasar spectra.
The HI surface density was determined by matching 21 cm emission with the same velocity line profile as the $\Hmol$ gas.
In order to mimic this method, we post-processed the galaxy to create velocity cubes of the $\Hmol$ and HI gas with a 2.5 pc by 2.5 pc by 10 km/s grid.
The velocity resolution was chosen to contain the typical line width of a MC.
For each cube in the grid, we then determined $X_{\Hmol}$ and $N_{\mathrm{HI}} + 2N_{\mathrm{H_2}}$.
The correspondence between our determination of the surface density and the observational data, particularly for large fractions of $\Hmol$, supports our use of $h$ for the column length.

\subsection{Resolution}
\label{sec:res}
%--------------------- Importance of resolution -----------
Over what resolutions does this implementation of $\Hmol$ hold and over what resolutions are high-density regions sufficiently well resolved to use it?
When simulating $\Hmol$, the mass and spatial resolution has the potential to affect both the densities of the gas particles and their smoothing lengths, $h$.
Molecular clouds have typical densities of 100 amu/cc, so the density structure within the simulations must be sufficiently resolved for gas with such densities to exist.
The resolution of the density structure is a function of both the mass resolution (number of particles) and the force resolution (the softening length and smoothing length).

%------------------- smoothing/softening -------------
Since the column densities used when calculating shielding are a function of the SPH smoothing, $h$, any resolution effect on them can also affect the fraction of $\Hmol$.
The value of $h$ is primarily a function of the particle number but it has a minimum value of 0.1 times the softening length.
In all of our simulations, we ensured that the minimum smoothing length was small enough that giant MCs could be resolved.

\begin{figure}
\centering
\includegraphics[width = 0.5\textwidth]{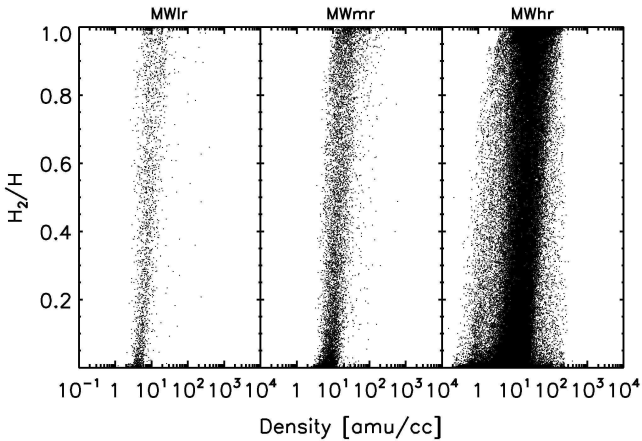}
\caption[Effect of mass resolution on $\Hmol$ abundance]
{The transition from atomic to molecular hydrogen for three isolated, Milky Way-like simulations with different mass resolutions.  The points represent individual SPH gas particles. For each of the simulations, the transition occurs at the same density, indicating that within this range of resolution the transition is independent of the number of particles.} 
\label{fig:resabund}
\end{figure}

%----------------------- data ------------
In order to determine the effect of mass resolution on the amount of  $\Hmol$ formed, we simulated the same disk galaxy with $10^5$ (MWlr), $10^6$ (MWmr) and $10^7$ (MWhr) gas particles.
The initial gas particle masses for each of the simulations were, therefore, of $1.4\times10^6$, $1.4\times10^5$ and $1.4\times10^4 \Msun$.
These mass resolutions were previously found to be sufficient to model the star formation and density structure in spiral galaxies prior to the addition of $\Hmol$ \citep{Christensen10}.
The effect of mass resolution on the fraction of $\Hmol$ as a function of density is shown in Figure~\ref{fig:resabund}.
We found that the relationship between density and the $\Hmol$ abundance was independent of the number of particles in the three simulations, indicating that the transition from HI to $\Hmol$ is consistent over this range of mass resolution.
This figure also shows that for our range of resolutions, the density structure is sufficiently resolved for gas with densities up to 1000 amu/cc to form.

\begin{figure}
\centering
\includegraphics[width = 0.5\textwidth]{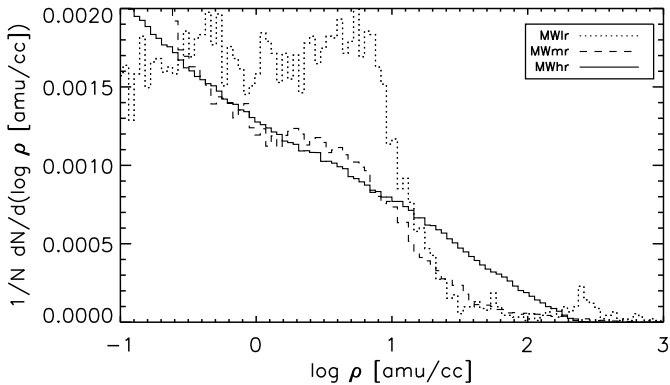}
\caption[Effect of mass resolution on gas density]
{Normalized histograms of the gas densities within the three isolated, Milky Way-like simulations with different mass resolutions. 
} 
\label{fig:resdens}
\end{figure}

However, the resolution does affect the fraction of gas that reaches high density (Figure~\ref{fig:resdens}).
As the resolution increases, the density distribution of gas particles becomes a smooth, declining function of density.
We, therefore, recommend particle masses of $1.4\times10^4  \Msun$ or less when using this method to simulate $\Hmol$.

\subsection{Star Formation}
%%Star Formation 
\label{sec:H2SF}
%--------------------- SF recipes
In our simulations with $\Hmol$ we incorporated a link between star formation and MCs by modifying the probability of a star forming to include the local fraction of $\Hmol$.
This link limits star formation to ``star-forming regions,'' represented by areas with abundant $\Hmol$.
We based our star formation recipe on the formalization of \citet{Stinson06}.
In this recipe star formation occurs stochastically, based on the local K-S relation in which the probability of star formation is a function of the local density.
Gas particles that are sufficiently dense ($\rho \geq \rho_{min}$) and cool ($T \leq T_{max}$) have a probability, $p$, of spawning a star particle.
The value of $p$ is a function of the local dynamical time $t_{form}$:
\begin{equation}
p = \frac{m_{gas}}{m_{star}}(1 - e^{-c^* \Delta t /t_{form}})
\end{equation}
where $m_{gas}$ is the mass of the gas particle, $m_{star}$ is the initial mass of the potential star particle, and $c^*$ is a star forming efficiency factor.

In order to incorporate $\Hmol$, we modified $c^*$ such that
\begin{equation}
c^*  = c^*_0~X_{\Hmol} 
\end{equation}
With this modification, the probability of star formation increases in highly molecular regions.
In our simulation without $\Hmol$, $c^* = c^*_0$ = 0.1, whereas with $\Hmol$, $c^*_0$ = 0.1.

In addition to incorporating $X_{\Hmol}$ into the star formation efficiency, we also altered the qualifications for a gas particle to form stars.
In our non-$\Hmol$ star formation recipe, $\rho_{min} = 100$ amu/cc, in order to mimic star formation in MCs.
In our $\Hmol$ star formation recipe, the dependence of $c^*$ on the molecular fraction already limits star formation to dense, molecular gas.
We, therefore, reduced $\rho_{min}$ to 0.1 amu/cc, which is sufficiently low to have negligible effect on the star formation.
Similarly, we also lowered $T_{max}$.  
In simulations without ${\Hmol}$, low-metallicity gas cannot cool below 10000 K so we set $T_{max} = 20000$ in order to select for the gas that would be in the warm and cold ISM.
When ${\Hmol}$ is included, however, a cold phase of the ISM is able to form.
In order to select for gas that had cooled, we set $T_{max} = 1000$ but found the star formation to be largely insensitive to this value.
The values for these star formation parameters are also outlined in Table 2.

When calculating star formation, the ability of the simulation to resolve the Jeans length and mass can be a factor.
\citet{BateANDBurkert97} showed that when the Jeans mass or and length of a particle exceeds the particle mass and smoothing or softening length, respectively, star formation becomes dependent on the specifics of the star formation recipe.
We chose not to include an artificial pressure floor to support the gas against gravitational collapse as other authors have done \citep[e.g.][]{RobertsonKravtsov08}.
Our simulations are sufficiently high resolution that gas particles for which the Jeans mass or length are unresolved necessarily meet our star formation criteria (including $\Hmol$ abundance).
Given  $c^*_0$ = 0.1, these particles will soon form stars, regardless of whether or not a pressure floor is included.
Therefore, Jeans instability below the resolution of our simulations is included within our sub-grid star formation model.
We found in several test simulations that the inclusion of a pressure floor made little difference to the total number or distribution of stars formed.

We acknowledge that as star formation in our simulations may happen below the resolution limit, artificial fragmentation can occur prior to star formation.
These simulations are, therefore, not suitable for studying such small scale phenomena as the spatial distribution of stars within star forming regions or the mass function of molecular clouds.
Instead, we focus on the total star formation and distribution of star formation above our resolution limit.

\subsection{Cosmological Simulations}\label{sec:cosmoIC}
We used cosmological simulations of a dwarf galaxy simulated with and without $\Hmol$ to study the effect of $\Hmol$ on the star formation and structure of the stellar disk.
Cosmological simulations of galaxies are necessary to study $\Hmol$ throughout galaxy evolution and in galaxies with realistically built structure.  
Dwarf galaxies were chosen as an extreme environment for studying gas dynamics, star formation, and the effect of SN feedback.
Here we describe the cosmological simulations presented in this paper.

We simulated the formation a field dwarf galaxy with an approximate final virial mass of $4 \times 10^{10} \mathrm{M_{\odot}}$ with $\Hmol$ (DH2) and compared it to the same galaxy simulated without $\Hmol$ (DnoH2).
We also included two lower resolution versions of the dwarf galaxy with $\Hmol$ (DH2\_mr and DH2\_lr).
While we have verified that our $\Hmol$ formation recipe holds across the range of resolutions of these simulations, star formation and feedback are also sensitive to resolution and dwarf galaxies may be particularly sensitive to changes in feedback.
We, therefore, included the lower resolution simulations to ensure that star formation converges at this resolution for this mass of galaxy when $\Hmol$ is included.
Aside from differences in resolution and as to whether or not $\Hmol$ was included, all the simulations are identical.
The simulations include cooling from H and He ions and from metal lines, cosmic UV background radiation, and a blastwave model for SN feedback, as described in section $\S$\ref{sec:general}.

The initial conditions for this dwarf galaxy were first used in \citet{Governato10} as dwarf galaxy 1 (DG1).
In that paper, it was simulated with neither full metal line cooling nor $\Hmol$ and produced a bulgeless dwarf galaxy.
The initial conditions consisted of a 25 Mpc box surrounding a halo selected from a low-resolution, DM only simulation.
The initial power spectrum used for the linear density field was calculated using CMBFAST code.
We assumed a concordance, flat, $\Lambda$-dominated cosmology with values from WMAP3 \citep{Spergel03}.

\begin{table*}
\begin{minipage}{165mm}
\caption[Cosmological Simulations Parameters for the Dwarf Galaxy]
{The cosmological simulations of the dwarf galaxy, listed with their resolution and star formation parameters.  
The two simulations listed in bold are the high-resolution runs and used in the plots throughout the remainder of the paper.
The average smoothing length of the gas particles, $\langle h\rangle$, is shown here for the disk gas, which was defined as having a density of 1 amu/cc or greater.
All simulations use the same initial conditions and include cooling from H and He ions and from metal lines, a cosmic UV background radiation, and a blastwave model for SN feedback.}
\label{tab:cosmoSims_H2}
\begin{center}
\begin{tabular}{lccccccccr}
\hline \hline
\                 &                  & \multicolumn{3}{c}{Particle Mass [$\Msun$]} &                   &                  & \multicolumn{3}{c}{Star Formation Parameters}\\
\cline{3-5} \cline{8-10} \\ 
                Name &                 $\Hmol$ &                 Dark &                 Star  &                 Gas &                 Softening    &                 $\langle h\rangle$  of Disk         &                 $c^*$ &                 $\rho_{min}$  &                 $T_{max}$   \\
                            &                                 &                           &                          &                          &                 Length [pc] &                 Gas [pc] &                            &                 [amu/cc]          &                 [K] \\ \hline
\bf{DnoH2}          & no                             & 16,000                & 1000                  & 3300                  & 87                                   & 60                                                        &   0.1                                          &     100            & 2e4 \\ 
\bf{DH2}               & yes                            & 16,000                &  1000                 & 3300                  & 87                                  & 60                                                         &    0.1 $X_{\mathrm{H_2}}$   &   0.1               & 1e3\\
DH2\_mr                & yes                            & 37,000                 &  2400                & 5900                  & 115                                 & 73                                                        &    0.1 $X_{\mathrm{H_2}}$   &   0.1               & 1e3\\
DH2\_lr                & yes                            & 126,000                 &  8000                & 20000                  & 173                                 & 100                                                        &    0.1 $X_{\mathrm{H_2}}$   &   0.1               & 1e3
\end{tabular}
\end{center}
\end{minipage}
\end{table*}

The two high-resolution simulations (DH2 and DnoH2) differ both in terms of whether $\Hmol$ was included and what star formation recipe was used.
In order to study the effect of limiting star formation to regions with $\Hmol$, we 1) multiplied the star formation efficiency, $c^*$, by the fraction of $\Hmol$ and 2) lowered both the threshold density and temperature, as previously described in \S~\ref{sec:H2SF}.
Aside from these changes, our parameters for star formation and feedback were the same as in \citet{Governato10}.
We assumed a \citet{Kroupa93} IMF and calculated SN energy using the scaling factor $E_{SN} = 10^{51}$ ergs per SN. 
The star formation parameters and information on resolution for each of the simulations can be found in Table~\ref{tab:cosmoSims_H2}.

%----------------------------------------------
\section[Results: Comparisons Between a Dwarf Galaxy Simulated with and without Molecular Hydrogen]{Results}  \label{sec:datasec}
We analyzed the effects of including $\Hmol$ and $\Hmol$-based star formation by comparing the total amount, distribution, and history of the gas and stars in our simulations.
In order to compare our simulations to observed galaxies, we generated mock-observations of our galaxies in multiple bands using SUNRISE \citep{Jonsson06}.
SUNRISE is a ray-tracing radiative transfer program that includes dust scattering and absorption.
We used it to measure photometric colors, luminosities, and star formation indicators such as H$\alpha$, FUV, and 24 $\mu$m.
In addition to simulating the stellar emission from the galaxy, we generated HI velocity cubes.
These velocity cubes were used for calculating the line profile when determining the galactic rotation speed and for determining the HI surface density when calculating the resolved and global K-S relations.
We produced the velocity cubes by using the smoothing kernel to distribute the gas particles spatially.
We then calculated the expected amount of 21cm emission from HI and binned it in x, y, and velocity space.
In order to make these results comparable to observations, we generated the velocity cube at a resolution and sensitivity typical of THINGS \citep{Walter08} for a dwarf galaxy at a distance of 5 Mpc.
As was done for the dwarf galaxies in THINGS, we generated data in 128 velocity channels with 1.3 km/s bins and a 1.5" pixel size. 
We then spatially smoothed the data cubes using a Gaussian beam with a full width at half max of 10"x10"
Finally, we made a sensitivity cut and discarded all emission from cells below a 2$\sigma$ noise limit, in which $\sigma = 0.65$ milli-Jansky/beam.

The final global properties for the galaxies simulated with $\Hmol$ (DH2, DH2\_mr, and DH2\_lr) and without $\Hmol$ (DnoH2) are listed in Table~\ref{tab:cosmoSimsFinal}.
These properties include the virial mass, stellar mass, mass of gas within $R_{vir}$, and mass of disk gas.
The disk gas for this purpose was defined as being the mass of HI plus the mass of $\Hmol$ (if applicable) multiplied by a factor of 1.4 to represent the presence of cold He.
The values of the Johnson B-band magnitude, $r_{25}$, and B-V color were calculated from SUNRISE observations.
The average and standard deviation of the SFRs were calculated over the last 2 Gyrs of the simulation.
Finally, the average metallicity of the gas (12 + log(O/H)) was calculated for the star forming gas, as further described in \S~\ref{sec:BTF}.

\subsection{Resolution}
From Table 2, it is apparent that the medium resolution simulation, DH2\_mr, was able to reproduce the star formation of DH2, including the total stellar mass, the stellar distribution ($r_{25}$), and the current SFR (B-V color and $\langle\mathrm{SFR}\rangle_{\mathrm{2 Myr}}$).
DH2\_lr was a less close match to DH2 and produced a slightly brighter and larger galaxy with approximately 1.16 times the total mass of stars.
We, therefore, conclude that star formation in this galaxy converges for resolutions of DH2\_mr and greater.
The initial masses of the gas particles in DH2\_lr and DH2\_mr are $2 \times 10^4 \Msun$ and $5900 \Msun$, respectively.
The convergence of star formation between these two mass resolutions is consistent with the recommendation from the isolated Milky Way-like simulations for gas particle masses of $1.4 \times 10^4$ or less (\S~\ref{sec:res}).
The total gas mass in the galaxies is more sensitive to resolution and there is an approximately 12\% difference in the gas mass within the virial radius between DH2\_mr and DH2\_hr.
This change indicates a slight decrease in the amount of gas within the halo with increasing resolution.

\begin{table*}
\begin{minipage}{165mm}
\caption{Final properties of the dwarf galaxy simulated with and without $\Hmol$.  
The two simulations listed in bold are the high resolution simulations and are used in the plots throughout the remainder of the paper.
Disk gas mass was defined as 1.4(HI + $\Hmol$), as is typically done in observations to account for the mass of cold He.  The average and standard deviation of the SFR were calculated over the last 2 Myr of the simulations, using bins of $5\times10^7$ years.  The metallicity, 12 + log(O/H), was calculated for the star forming gas, as described in \ref{sec:metallicity}.}
\label{tab:cosmoSimsFinal}
\begin{center}
\begin{tabular}{lcccccccccr}
\hline \hline
                	& \multicolumn{4}{c}{ Mass  [$10^8 \Msun$]	} &                 &                 &                 \\
\cline{2-5} \\ 
Name 			& Virial 	&  Stellar 		&                Gas			  	&                Gas 	&               $\mathrm{M}_{\mathrm{B}}$	&               $r_{25}$	&               B-V	&                $\langle \mathrm{SFR}\rangle$	&                 $\sigma_{\mathrm{SFR}}$	& 12 + log(O/H)                \\
               			&                	&               		&               $(R < R_{vir})$ 		&                	(Disk)&              	&      [kpc]         	&			&                $[\Msun yr^{-1}]$					&                $[M_{\sun} yr^{-1}]$ 			&               \\ \hline
{\bf DnoH2}		&   $360$	&   $2.1$ 		& $17$  				& $4.3$ 		& -15.27			& 1.13			& 0.49		&  0.0018  								& 0.0041 							& 7.71 \\ 
{\bf DH2}            	&   $380$ 	&   $2.5$ 		& $23$ 				& $5.5$ 		& -15.70			& 1.96			& 0.38		& 0.0073	 								& 0.0018 							&  7.90 \\ 
DH2\_mr       	  	&   $390$ 	&   $2.4$ 		& $26$  				& $6.1$ 		& -15.73			& 2.00			& 0.36		& 0.0077 									&  0.0027							&  7.53 \\	 
DH2\_lr       	  	&   $390$ 	&   $2.9$ 		& $29$  				& $8.0$ 		& -16.37			& 2.79			& 0.26		& 0.0160 									&  0.0045 							&  7.81 
\end{tabular}
\end{center}
\end{minipage}
\end{table*}

\subsection{Stellar and Baryonic Mass}
\label{sec:BTF}
%Baryonic Tully-Fisher
In order to compare the global properties of the simulations, we first studied the total amount of stars and gas contained in a DM halo.
In dwarf galaxies, being able to correctly simulate the luminosities of galaxies as a function of their DM halo mass is key to the missing satellite debate \citep{Moore99}.
This debate addresses why simple extrapolations of the luminosity function from DM-only simulations predict more low-mass satellites than are observed.
The luminosity of the galaxy and the amount of baryons contained in a DM halo are a function of the efficiency of star formation, the amount of gas that is accreted, and the amount of gas that is lost as the result of feedback.
The inclusion of $\Hmol$ has the potential to affect the amount of gas that turns into stars and the amount of SN feedback.
The efficiency of SN feedback is affected by the state of gas, with clumpier gas leading to greater gas loss.
Therefore, the inclusion of $\Hmol$ could change the amount of gas lost by producing a denser, colder ISM.

In order to measure the effect on the stellar and baryonic mass, we determined the dark, stellar, and gas mass  of the galaxies (Table~\ref{tab:cosmoSimsFinal}) and calculated their positions on both the standard and baryonic Tully-Fisher relationships.
While the standard Tully-Fisher relationship relates the luminosity of galaxies to their rotation velocity, the baryonic Tully-Fisher relationship relates the amount of baryonic mass in a halo to its rotation velocity.
The baryonic Tully-Fisher relationship is a particularly important relationship for low-mass galaxies because of the wide spread in their stellar-to-mass ratios.
It has been observed to be an exceptionally tight relationship \citep{McGaugh05} and can, therefore, provide a stronger constraint than the standard Tully-Fisher relationship.

\begin{figure}
\centering
\includegraphics[width = 0.5\textwidth]{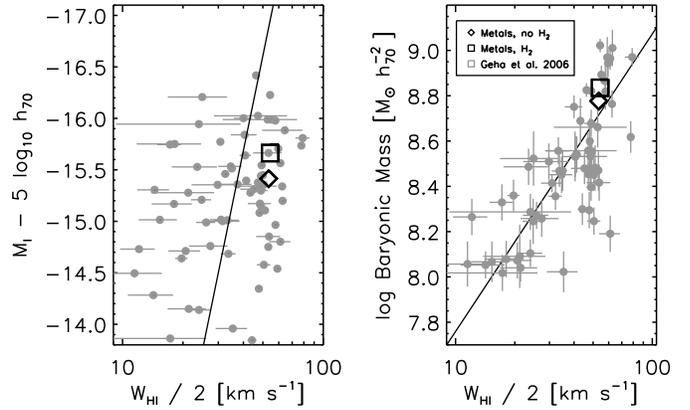}
\caption[Tully-Fisher relationship for dwarf galaxies with and without $\Hmol$]
{Standard (left) and baryonic Tully-Fisher (right) relations for the two simulations, overlaid on observational data from  \citet{Geha06} for dwarf galaxies (grey filled circles).  
The diamond represents DnoH2 and the square represents DH2.   
Both simulations lie within the observed scatter at a redshift of zero but the inclusion of $\Hmol$ resulted in a slightly higher stellar and baryonic mass.
} 
\label{fig:BTF}
\end{figure}

In Figure~\ref{fig:BTF}, we compare the luminosities and baryonic fractions of our galaxies to the observational data compiled in Figure 6 of \citet{Geha06}.
For both relationships, we measured the HI line width at the 50\% level ($\mathrm{W_{21}}$) as in \citet{Haynes99}, using the 21cm line from a simulated HI velocity cube with the galaxy oriented at $45^\circ$.
When comparing our galaxies to the standard Tully-Fisher relationship, the I-band magnitude was calculated from a SUNRISE simulated observation.
When comparing the galaxies to the baryonic Tully-Fisher relationship, we followed the same method as \citet{Geha06} to find the baryonic mass: the total gas mass was assumed to be 1.4 times the HI mass and the stellar mass was determined from the SDSS {\em i}-band magnitudes and {\em g} - {\em r} colors (calculated with SUNRISE) and the \citet{Bell03} mass-to-light ratios.
Both data points are within the observational spread, although the presence of $\Hmol$ resulted in a galaxy richer in both stars and cold gas.

As seen by the positions of the galaxies in the Baryonic Tully-Fisher and quantified in Table~\ref{tab:cosmoSimsFinal}, the inclusion of $\Hmol$ increased the baryonic mass to 1.28 times the baryonic mass in DnoH2.  
Most of this increase was in the form of gas: the mass within $R_{vir}$ for DH2 was 1.29 times higher than in DH2.
The stellar mass increased to only 1.16 times higher than in DnoH2, indicating a slightly lower stellar mass to gas mass fraction.

We investigated the source of DH2's larger baryonic fraction by finding the total mass of gas ever in the galaxies and ever ejected by SNe from the galaxies.
We calculated the mass of gas that was ever in the galaxy by identifying the all gas particles that were included within the main halo at any timestep.
We then found the mass of gas lost from the galaxy by identifying all gas particles that had ever been heated by SNe and that were once part of the main halo but were outside the halo at z = 0.
Our method for calculating the mass of gas ever in the galaxy does not include gas that became stars in separate smaller halos prior to their accretion.
It also has the potential to miss gas that was accreted onto and expelled from the halo between two adjacent timesteps.
However, we are interested in the comparison between the two simulations rather than the absolute amounts so these systematic uncertainties are relatively unimportant.

We found little difference in the total mass of gas accreted onto DnoH2 and DH2.
DnoH2 and DH2 had a total masses of gas ever in the galaxy of $3.71 \times 10^9 \Msun$ and $3.64 \times 10^9 \Msun$, respectively
There was actually slight {\em increase} in the amount of gas expelled by SNe beyond the virial radius in DH2 compared to DnoH2.
In DH2, a total of $5.4 \times 10^8 \Msun$ was expelled by SNe, compared to $3.9\times10^8 \Msun$ expelled from DnoH2.
The decreased baryonic mass of DnoH2 compared to DH2 was, therefore, not the result of either lower amounts of gas accretion or greater amounts of SNe-driven gas loss.
Instead, the greater amounts of gas loss in DnoH2 came from gas being smoothly stripped from the outskirts of the halo.
This stripping of gas may be an indication that gas in DnoH2 was less tightly bound than in DH2 or it may be the result of stochastic variation in the mergers.

\subsection{Metallicity}
\label{sec:metallicity}
\begin{figure}
\centering
\includegraphics[width = 0.5\textwidth]{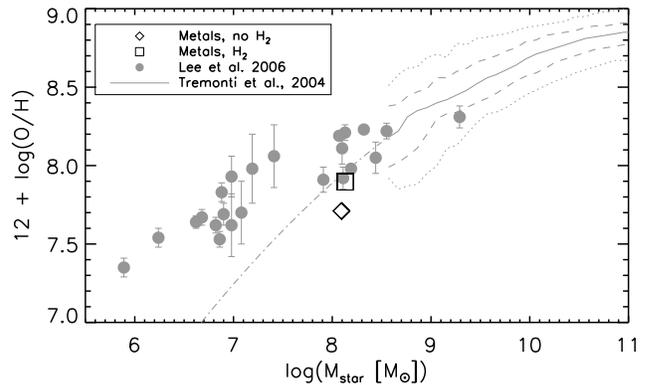}
\caption[Mass-metallicity relationship.]
{Mass-metallicity relationship with DnoH2 (diamond) and DH2 (square) overlaid on observational data.
The stellar masses of the simulated galaxies were calculated in an observationally-motivated fashion from the Johnson-Morgan K-band luminosity.
The metallicities shown for the simulated galaxies are the mean metallicities of the gas particles scaled by the probability of star formation.
The filled grey circles represent observed dwarf irregular galaxies from \citet{Lee06}.
The grey lines are the observational fits from \citet{tremonti04}: the solid line is the best fit line, the dot-dashed line is that fit extended to lower stellar masses, and the dashed and dotted lines show the 1 and 2 $\sigma$ spread, respectively.
The metallicity of DH2 is consistent with the observed galaxies.
The metallicity of DnoH2 is slightly lower than the observed galaxies but may be consistent with the spread in metallicities determined in \citet{tremonti04}.
} 
\label{fig:mz}
\end{figure}

The total stellar mass, in addition to the amount of gas lost and accreted, is responsible for the metallicity of the galaxy.
Given the importance of metallicity to the formation and shielding of $\Hmol$, we show the simulated galaxies on the observed mass-metallicity relationship (Figure~\ref{fig:mz}).
In this figure, the observational data is taken from the \citet{Lee06} set of dwarf irregular galaxies and the \citet{tremonti04} fits for their survey of SDSS galaxies.
The latter is scaled down by 0.26 dex to reflect the tendency of the method used to overestimate the oxygen abundance relative to other methods \citep{tremonti04,Erb06, Brooks07}.
DH2 lies on top of an observed galaxy and very close to the \citet{tremonti04} fit.
The metallicity of DnoH2 is slightly lower than that of observed galaxies of the same stellar mass but may lie within the spread predicted from  \citet{tremonti04}.

The stellar masses of the simulated galaxies used in the mass-metallicity relationship were calculated from mock-observations of the stellar emission using the same scaling as in \citet{Lee06}.
The mass-to-light ratio was determined from the B-K color and combined with the magnitude in the K-band to produce a stellar mass.
The metallicities in this plot were determined for the star forming gas.  
In order to account for the fact that observations of gas metallicity are determined for HII regions, the metallicities of the gas were scaled by the star formation probability, as in \citet{Dave07} and \citet{Finlator08}.
Compared to the average metallicity of all HI and $\Hmol$ gas within the radius where star formation has occurred in the last 100 Myrs, this scaling has the potential to change the metallicity by up to 0.1 dex.
In both these galaxies a substantial amount of low-metallicity HI  gas lies in an extended disk just beyond the star formation radius.
Therefore, while the star formation takes place in primarily higher metallicity gas, extremely low-metallicity gas chemistry remains important to the galaxy at z = 0. 

%%%%%%%%%%%%%% Commented Out %%%%%%%%%%%%%%%%%

\subsection{Star Formation Histories}
\label{sec:SFH}
In order to determine what circumstances led to differences in the stellar masses of the two simulations, we compared the evolution of the star formation histories of DH2 and DnoH2.
Previous authors \citep[e.g.][]{RobertsonKravtsov08,Gnedin10,Krumholz11a} have suggested that introducing a metallicity dependency to star formation by linking star formation to $\Hmol$ could suppress star formation at high redshift.
Using galaxy evolution simulations with $\Hmol$-based star formation, \citet{Gnedin10} and \citet{Kuhlen11} hove show reduced star formation up to a redshift of 3 and 4, respectively.

\begin{figure}
\centering
\includegraphics[width = .5\textwidth]{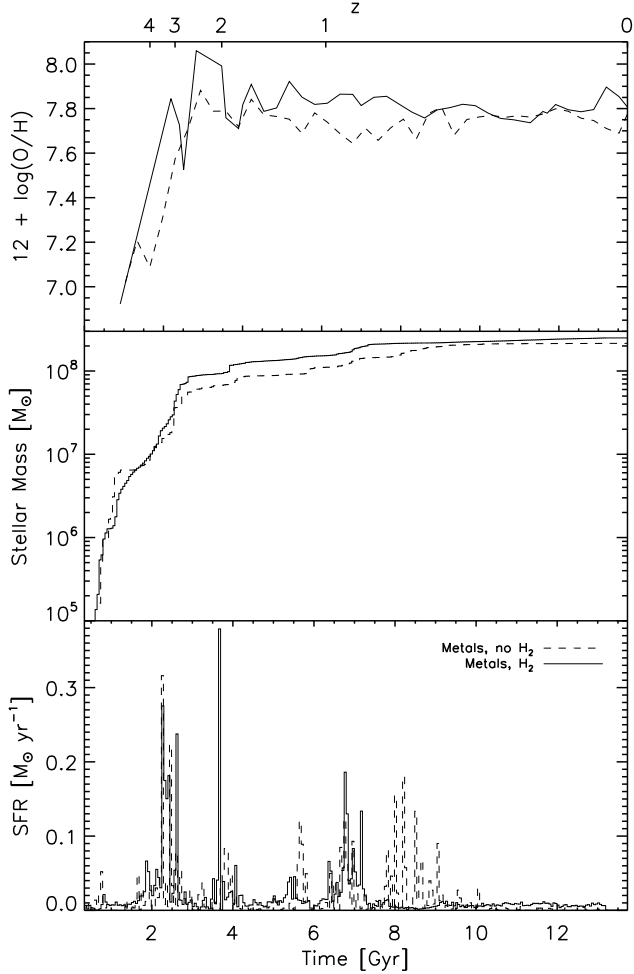}
\caption{The evolutions of the gas metallicity and stellar mass, and the star formation histories.
In each of the panels, the dashed line represents DnoH2 and the solid line represents DH2.
The top panel shows the mean gas-phase metallicity for the largest progenitor of final galaxy.
The middle and bottom panels show the star formation histories.
In these two plots, all star particles within the galaxy at z = 0 are included.
The middle panel shows the cumulative star formation histories (stellar mass formed as a function of time).
The bottom panel shows the SFRs over time.
There is a very slight decrease in star formation in DH2 prior to z = 4.
In general, though, the star formation rates track together with the slightly higher rates in DH2 reflecting the greater disk mass.}
\label{fig:SFH}
\end{figure}

We examine whether low-metallicities at high redshift may have suppressed star formation in DH2 relative to DnoH2 by comparing the evolution of the gas-phase metallicity to the SFR and total stellar mass (Figure~\ref{fig:SFH}).
The top panel shows the mean gas-phase metallicity of the star forming gas in the largest progenitor as a function of time.
Notably, there is almost no evolution in the gas-phase metallicity since a redshift of four.
For the majority of the galaxies's histories, the inflow of metal-poor gas has balanced the production of metals from stars.
The second and third panels show the cumulative star formation histories and the star formation histories  of the galaxies.
Unlike the metallicity panel, these panels show the history of all star particles that are in the galaxy at $z = 0$, rather than only those particles that were in the galaxy at a given redshift. 

Figure~\ref{fig:SFH} shows slight evidence for a decrease in high redshift star formation.
DH2 does initially have a slightly lower stellar mass than DnoH2 but by $z = 4$, DH2 has begun to surpass the stellar mass of the other galaxy.
At this same redshift, the metallicity of the star forming gas settles to a near constant state.
With our limited number of high redshift outputs, it is not clear if the decreased stellar mass at $z > 4$ is the result of a higher star formation threshold or simply stochastic variation in the merger history.
Regardless, any suppression of star formation compared to DnoH2 is of brief duration.

In general, star formation in our simulations appears very robust to the specifics of the star formation recipe.
Essentially, if the star formation efficiency is decreased either artificially or because of decreased $\Hmol$ content, the star formation may be delayed but the particle will continue to grow denser and more molecular and will eventually form a star.
For these simulations, SN feedback is the primary regulator of star formation.
The relative importance of $\Hmol$ formation and SN feedback and how it compares across different simulations is addressed in further detail in the discussion.

\subsection{Color and Extent of Stellar Disk}
\label{sec:prof}
\begin{figure}
\centering
\includegraphics[width = 0.5\textwidth]{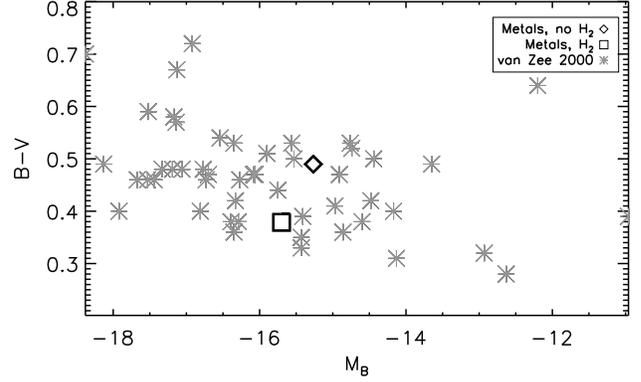}
\caption[Colors and magnitudes for dwarf galaxies with and without $\Hmol$]
{B-V color versus B-band magnitude for the simulated galaxies overlaid on observations of field dwarfs from \citet{Zee00} (grey asterisk).  
The diamond represents DnoH2 and the square represents DH2.
The inclusion of $\Hmol$ resulted in the dwarf galaxy being brighter and bluer, indicating a greater amount of on-going star formation.} 
\label{fig:color}
\end{figure}

In addition to changes in the total mass, we examined the effect of $\Hmol$ on the current star formation.
From the data in Table~\ref{tab:cosmoSimsFinal}, the SFR over the last two Myrs was four times higher in DH2 than in DnoH2.
This increase is apparent from an observational standpoint in the colors of the galaxies.
In Figure~\ref{fig:color}, we compared the B-band magnitudes and B-V colors of the galaxies to observational data for dwarf galaxies from \citet{Zee00}.
While both the colors and magnitudes of DH2 and DnoH2 are consistent with the observed data, we found that the addition of $\Hmol$ resulted in brighter and bluer galaxies.
The brighter B-band magnitudes of DH2 were the result of both a larger stellar mass ($\S$~\ref{sec:BTF}) and more recent star formation.

\begin{figure}
\centering
\includegraphics[width = 0.5\textwidth]{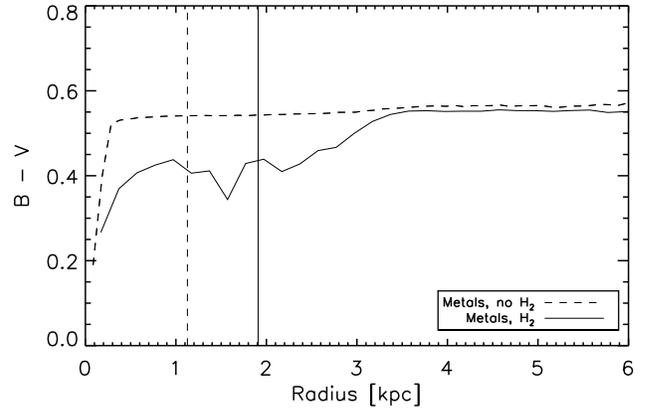}
\caption[Colors of galaxies with and without $\Hmol$ as a function of radius]
{Color as a function of radius for DnoH2 (dashed line) and DH2 (solid line).  The vertical lines represent $r_{25}$ for each galaxy.  
The bluer colors at the optical radii in DH2 indicate on-going star formation throughout a larger area of the disk.} 
\label{fig:colorProf}
\end{figure}

In order to compare the distribution of stars and star formation in the two galaxies at z = 0, we used SUNRISE to determine the radially-binned B-V colors for the galaxies when oriented face-on (Figure~\ref{fig:colorProf}).
Overlaid on the plot are vertical lines representing the locations of $r_{25}$.
DH2 had much more spatially extended star formation, as can seen by the blue colors that extend out to several kpc in radii.
This extended star formation is evident in a larger stellar disk, characterized by the greater value of $r_{25}$.
This distribution of colors is consistent with observations of the color profiles of dwarf irregular galaxies in \citet{Zee00}.
In contrast, DnoH2 is distinctly blue at the center and surrounded by a dim, redder population.
This color profile indicates that the small amount of star formation at z = 0 was extremely concentrated at the center of the galaxy.
Therefore, the addition of $\Hmol$ resulted not only in galaxies with greater amounts of star formation at z = 0, but also caused star formation to occur over a larger extent.

\subsection{Kennicutt-Schmidt Relation}
\label{sec:KS}
Based on the changes to the star formation caused by the inclusion of $\Hmol$, it is important to compare the star formation laws in both simulated galaxies to the observed K-S relation.
The K-S relation has been observed both as the global K-S relation on galaxy wide scales \citep{Kennicutt98a} and as the resolved K-S relation on hundred-parsec scales \citep{Bigiel08}.
The existence of a global K-S relation implies a consistent relationship between star formation and gas surface density when averaged over the entire disk of the galaxy.
On the sub-kpc scale, the correlation requires this relationship hold even as smaller star-forming regions begin to be resolved.

Star formation recipes in simulations that are based on the free-fall time, as ours is, imply a K-S relation for gas above the star formation density threshold ($\rho_{min}$).
Such recipes that allow star formation in all disk gas by having a low $\rho_{min}$, such as 0.1 amu/cc, are more directly related to the global K-S relation.
In contrast, star formation recipes, such as ours, which limit star formation to higher density regions through the use of high-density thresholds or $\Hmol$-based star formation are more directly related to the resolved K-S relation.
Resolving high-density regions in simulations and tying star formation to those regions through the use of a high density threshold improves the spatial distribution of stars \citep{Saitoh08}.
The use of a high-density threshold, however, removes the direct connection between the SFR and the global gas distribution.
Fitting the global K-S relation requires that the density structure of the gas result in the appropriate fraction of high density, star-forming gas.
In order to verify that star formation in our galaxies is occurring at the correct rate over all spatial scales, we replicated both the resolved K-S relation and the global K-S relation.

In order to replicate the resolved K-S relation, we determined HI, $\Hmol$, and SFR surface densities in $750\pc \times 750\pc$ boxes within $r_{25}$ and compared them to the gas and SFR surface densities measured over the same spatial scale and within $r_{25}$ for dwarf galaxies \citep{Bigiel10}.
Our HI surface densities were determined from the zeroth moment of a simulated velocity cube of 21cm emission with average THINGS resolution for the galaxy oriented at a $45^{\circ}$ angle.
This velocity cube is described in detail at the beginning of $\S$3. 
We corrected for the inclination by multiplying all surface densities by sin($45^{\circ}$).
We calculated the SFRs as in \citet{Bigiel08}, from a combination of FUV and 24$\mu$m flux in a SUNRISE mock-observation.
The surface densities of $\Hmol$ were determined directly from the simulation, without reference to CO.

The observational K-S data we compared to is a subset of the data presented in \citet{Bigiel10}.
The K-S data presented in \citet{Bigiel10} is for all the THINGS galaxies, including the more metal-rich spiral galaxies and the low-metallicity dwarf galaxies.
Since the surface density at which HI transitions to $\Hmol$ is metallicity dependent, we limited our comparison to only the metal-poor dwarf galaxies from THINGS: DDO 53, DDO 153, Holmberg I, Holmberg II, IC 2574, M81 Dwarf A, and M81 Dwarf B.
These galaxies range in metallicity from $7.54 \leq$12 + [O/H] $\leq 7.94$ \citep{Moustakas06} and the majority of points come from IC 2574, which has a metallicity of 12 + [O/H] = 7.94, very similar to DH2.
The gas surface densities shown are equal to 1.4$\Sigma_{\mathrm{HI}}$; $\Hmol$ is not included in the observations because of the lack of observed CO and uncertainty in $X_{\mathrm{CO}}$. 
Low metallicity dwarf galaxies are HI dominated, however, so neglecting $\Hmol$ should have only a small effect on the surface density.
 \citet{Bolatto11} (not shown in the figure) is another recent observational study of the resolved K-S relation in low metallicity environments.
 In that study, very high resolution (~12 pc) measurements of the HI and total gas surface density were taken in order to calculate the $\Hmol$ surface density independent of CO emission.
 They observed similar values of $\Sigma_{\mathrm{SFR}}$ at ~3 times higher gas surface densities than the  \citet{Bigiel10}.
 This discrepancy may be due to physical differences between individual galaxies or the observational techniques used.
 
Both simulations reproduce the steep slope of the resolved K-S relation at low surface densities.
In DnoH2, the steep slope at lower surface densities was generated by the minimum density requirement for star formation.
In DH2, the steep slope marks the transition from atomic to molecular gas.
DH2 has a higher SFE than DnoH2, as evident by the lower surface densities at which star formation takes place.
The higher SFE in DH2 could be because the likelihood of star formation increased gradually with increasing $\Hmol$ fraction, rather than turning on sharply at the star formation threshold density, $\rho = 100$ amu.
It also could reflect greater heterogeneity in the gas density in DH2 (see discussion), such that the high density of the star-forming gas was tempered by low density nearby gas.

In addition to the slightly lower gas surface densities in DH2, the primary differences between DnoH2 and DH2 are in the spread and number of points.
In DH2, the greater amount of spread is likely because the density threshold for star formation in DnoH2 was replaced by the more gradual transition as the gas become molecular.
The greater number of data points for DH2 is a reflection of the greater area in the galaxy forming stars at a redshift of zero, as discussed in $\S$\ref{sec:prof}.

\begin{figure}
\centering
\includegraphics[width = .5\textwidth]{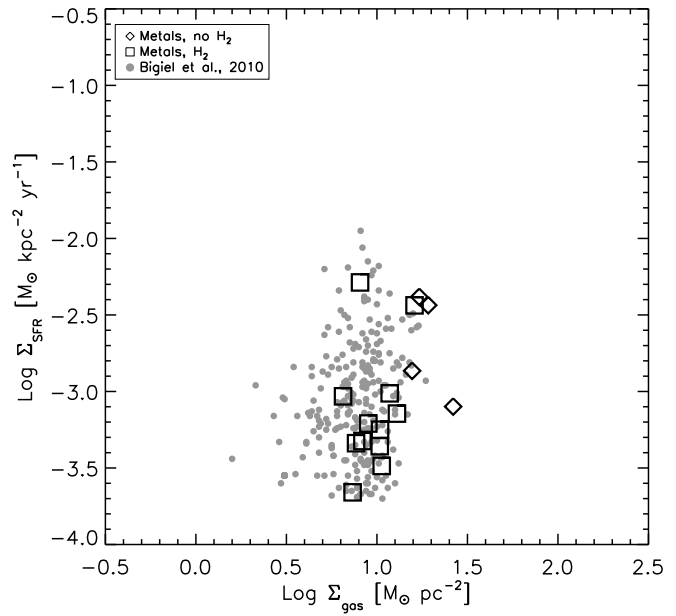}
\caption[The resolved Kennicutt-Schmidt relation for dwarf galaxies with and without $\Hmol$, calculated using simulated observations]
{The resolved K-S relation within $r_{25}$ for the simulations at z = 0.
The simulation data are overlaid on observational data of low metallicity dwarf galaxies from \citet{Bigiel10} and of the SMC \citet{Bolatto11}.  
The surface density of HI was computed from a mock-THINGS observation and the SFRs were determined from SUNRISE-generated mock observations of the FUV and 24$\mu$m flux.  
The diamonds are for DnoH2 while the squares are for DH2.  
Both simulations reproduce the steep slope at low surface densities.
DH2, however, has star formation at sightly lower surface densities, more star-forming gas, and a greater spread of data points.} 
\label{fig:KSres}
\end{figure}

\begin{figure}
\centering
\includegraphics[width = .5\textwidth]{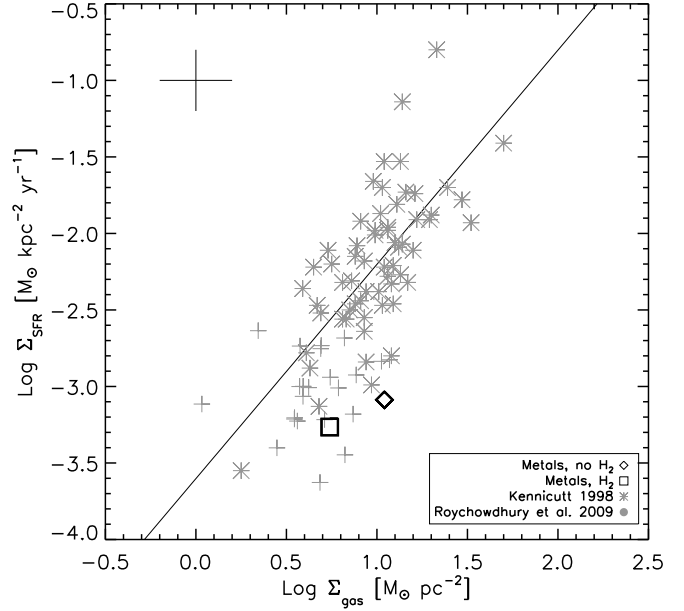}
\caption[The global Kennicutt-Schmidt relation for dwarf galaxies with and without $\Hmol$, calculated using simulated observations]
{The global K-S relation for the simulations at z = 0 overlaid on observational data from spiral galaxies \citep[asterisks;][]{Kennicutt98a} and dwarf galaxies \citep[filled circles;][]{Roychowdhury09}. 
The diamond represents DnoH2 and the square represents DH2.  
The cross at the top of the plot represents the average error for the observational data in \citep{Kennicutt98a}.  
The line is the K-S power-law with an exponent of 1.4 for spiral and starbursting galaxies.
DH2 fits well with the observed dwarf galaxies and DnoH2 lies very near to the observed galaxies.} 
\label{fig:KS}
\end{figure}

In order to compare DH2 and DnoH2 to the global K-S relation (Figure~\ref{fig:KS}), we used SUNRISE-generated H$\alpha$ emission to find the SFR at z=0.
We calculated the total HI and $\Hmol$ surface density within $r_{25}$, where $r_{25}$ was calculated from a Johnson B-band SUNRISE observation and HI and $\Hmol$ surface densities were determined as for the resolved K-S relation. 
We then compared the location of the simulated galaxies on the global K-S relation to both spiral \citep{Kennicutt98a} and dwarf \citep{Roychowdhury09} galaxies.
Despite the fact that the final SFR in DH2 was four times higher than in DnoH2 (Table~\ref{tab:cosmoSimsFinal}), both galaxies have similar locations on the K-S relation.
This is because the star formation in DH2 was much less spatially concentrated than in DnoH2.
As in the resolved K-S relation, DH2 fits particularly well with the observed galaxies while DnoH2 has a slightly higher gas surface density than observed galaxies with similar star formation surface densities.
Both simulated galaxies are below the K-S relation best-fit line and lie within the scatter of observed galaxies.

It is gratifying that both DH2 and DnoH2 fit more closely with the observed dwarf galaxies than spiral galaxies.
The lower ratios of $\Sigma_{\mathrm{SFR}}$ to $\Sigma_{\mathrm{gas}}$ in dwarf galaxies indicate smaller fractions of star-forming gas.
These lower amounts of star-forming gas are presumed to be the result of lower-metallicities, which raise the surface density where $\Hmol$ is found, and smaller fractions of gas that reach high surface densities.
While \citet{Roychowdhury09} does not contain metallicity measurements for these galaxies, they can be assumed to lie on the stellar mass-metallicity relationship.
With B-band magnitudes ranging from -14.90 to -10.78, they are even dimmer than our simulated galaxies ($M_{\mathrm{B, DH2}} = -15.70$ and $M_{\mathrm{B, DnoH2}} = -15.27$) and likely have metallicities that are similar or even smaller.
Both DH2 and DnoH2 have lower SFEs than would be predicted from the global K-S relation because 1) their high star formation thresholds limit the fraction of gas capable of forming stars and 2) these galaxies have fractions of star-forming gas more similar to observed dwarf galaxies than spiral galaxies.

In addition to calculating the location of the galaxies on the KS law, we calculated the molecular gas depletion time ($\Sigma_{\Hmol}/\Sigma_{\mathrm{SFR}}$) on 750 pc scales.
$\Hmol$ forms at the correct surface densities and the star formation efficiency has been tuned to produce appropriate amounts of star formation. 
However, the inability of the code to resolve the interior structure of molecular clouds in connection with stellar feedback results in artificially low amounts of $\Hmol$ near recent star formation.
Essentially, in our model it is impossible for $\Hmol$ to remain in a gas particle that has recently formed a star because of the spatial scale of SN feedback.
We account for this effect when calculating $\Sigma_{\Hmol}$ by adding in the amount of $\Hmol$ when star formation occurred.
Using this rough estimation, we find a smaller average molecular depletion time of $10^8$ years, compared to molecular depletion time on 100 parsec scales of $1.6 \times 10^9$ years for the SMC \citep{Bolatto11} and $2.0 \times 10^9$ years for the THINGS spiral galaxies \citep{Bigiel08}.
Note that the observed molecular depletion times are generally calculated for higher star formation rate surface densities: $10^{-3} \lesssim \Sigma_{\mathrm{SFR}} \lesssim 10^{-1}$, as opposed to our data which had $\Sigma_{\mathrm{SFR}} \leq 10^{-3}$.
The shorter depletion times of our simulations may stem from the uncertainty in our calculation and the different $\Sigma_{\mathrm{SFR}} $ of the observations it is being compared to or it may indicate a need to raise the clumping factor, $C_\rho$, used in the $\Hmol$ formation calculation.

The lack of contrast between DnoH2 and DH2 in either version of the K-S relation is notable.
Even with very different models for the interstellar media and for star formation, the results for both are similar to each other and to the observed data.
For these two simulations, the relationship between gas surface density and star formation is similar since both star formation recipes are based on the free-fall time and both impose a high-density threshold (in DnoH2, an explicit threshold, in DH2, an implicit one based on the $\Hmol$ abundance).
The similar results for DnoH2 and DH2 indicates that while the K-S relation is important to reproduce, it is an imprecise tool for distinguishing between these two simulations.
A similar correspondence between the K-S relation generated with a high density threshold for star formation and a $\Hmol$ dependency was presented in \citet{Kuhlen11}.

\section{Discussion}
\label{sec:disc_dwarf}
\subsection{Changes in Gas Thermodynamics}

The increase in baryonic mass and star formation at z = 0 is most likely the result of the increased amount of cold, dense gas formed in our simulation, caused by the inclusion of $\Hmol$ and the shielding and cooling that accompany it.
In incorporating $\Hmol$, we included shielding of HI by dust, in addition to dust and self-shielding of $\Hmol$.
This shielding of HI and $\Hmol$ gas from photoheating drastically reduces the heating rates of the dense disk gas.
In simulations that have both shielding and low temperature (T $\leq 10^4$) cooling, either from metal lines of $\Hmol$ (see the following paragraphs), the lack of heating will result in cold gas.
In DH2, both of these conditions are met, whereas in DnoH2, the gas was unshielded, in addition to having smaller cooling rates.

\begin{figure}
\centering
\includegraphics[width = 0.5\textwidth]{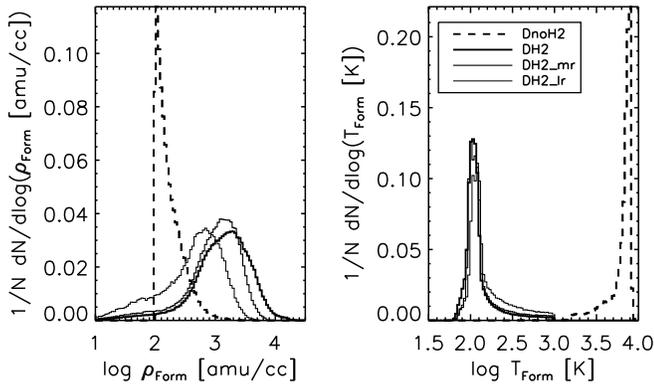}
\caption[Distribution of gas densities and temperatures prior to star formation]
{Normalized histogram of the densities (left) and temperatures (right) of the gas particles that formed stars. The dashed line represents DnoH2, while the solid lines are for simulations with $\Hmol$ at the three different resolutions, DH2\_lr (thin), DH2\_mr (medium weight), DH2 (thick).  The densities and temperatures of gas that form stars in simulations with $\Hmol$ are very similar across all resolutions.  Star formation in simulations with $\Hmol$ in higher density and colder gas than in DnoH2.  } 
\label{fig:sfgas}
\end{figure}

Evidence for the increased amount of cold, dense gas caused by our $\Hmol$ implementation is apparent when comparing the star-forming gas in DnoH2 and DH2.
Figure~\ref{fig:sfgas} contains histograms of the density and temperature of the gas from which stars formed in the simulations.
In simulations with $\Hmol$, the formation of a cold phase of the ISM is apparent in the colder and denser star-forming gas.
The star-forming gas in these simulations is much more similar to the star-forming gas in actual galaxies.

\begin{figure}
\centering
\includegraphics[width = 0.5\textwidth]{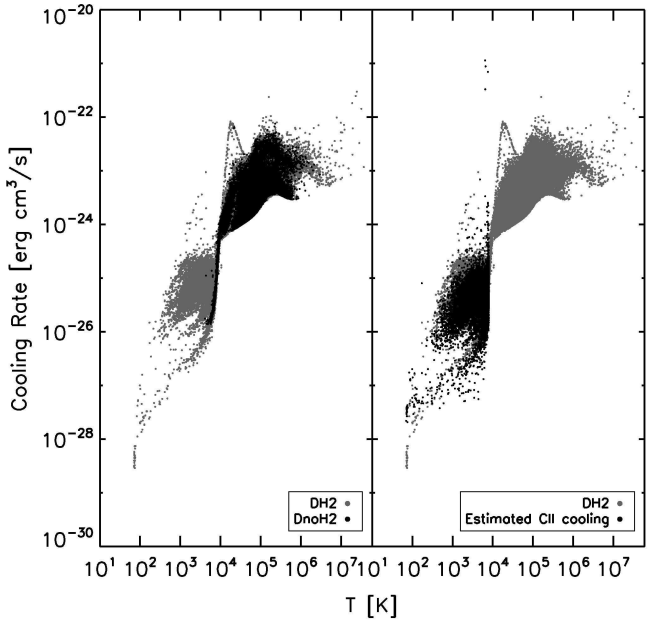}
\caption[Cooling rate of gas with and without $\Hmol$ as a function of temperature]
{Cooling rate for gas at a redshift of zero.  
The left panel compares the rates of gas cooling between DnoH2 (black) and DH2 (grey). 
The points represent individual SPH gas particles.
The metallicity of gas particles shown in this image covers the range $-1.5 \leq \mathrm{log}(Z/Z_{\sun}) \leq -0.5$.
When $\Hmol$ was included, the gas was able to cool to lower temperatures and had a higher cooling rate at temperatures below $10^4$ K. 
The right panel compares the rates of gas cooling in DH2 (grey) to the estimated maximum cooling rate from  CII, assuming all carbon is signally ionized.
The estimated CII cooling rate is similar to that from $\Hmol$ in low-metallicity regions with molecular gas.}
\label{fig:coolcurve}
\end{figure}

While the incorporation of shielding is primarily responsible for the increased amount of cold gas, $\Hmol$ also acts as an additional source of cooling at temperatures between 200 and 5000 K (Figure~\ref{fig:coolcurve}).
The importance of cooling from $\Hmol$ in relation to other sources of cooling was heightened in the dwarf galaxy in part because of its low metallicity.
The metallicity of the star forming gas in DH2 was 12 + log(O/H) = 7.9 and the metallicity of the disk gas beyond the star forming radius was even lower.
At these metallicities, the metal line cooling will be less effective than in spiral galaxies.

The relative importance of $\Hmol$ to metal line cooling in our simulations is also heightened by the different physics assumed when calculating the $\Hmol$ and the metal line cooling, leading to an underestimate of the metal line cooling rates compared to the $\Hmol$ cooling rates.
The calculation for the $\Hmol$ abundance includes a model for shielding of HI and $\Hmol$ and a model for dissociating radiation from the interstellar radiation field (ISRF).
The metal line cooling rates were introduced to GASOLINE in order to better model the enrichment and cooling of the intergalactic medium \citep{Shen10}.
These cooling rates were calculated assuming optically thin gas for a range of temperatures, densities, and redshift-dependent UV background radiation using CLOUDY \citep[version 07.02;][]{Ferland98}.
These assumptions are valid in the intergalactic medium.
However, in the ISM, the lack of UV radiation from the ISRF, even with the inclusion of cosmological UV background radiation, results in a much lower UV flux.
The lack of UV radiation decreases the rate of cooling from forbidden line transitions in unshielded gas, in particular CII.
With more forbidden line transitions, the cooling rates for the low temperature gas in DnoH2 would be expected to substantially increase.
The observed difference in cooling rates between DH2 and DnoH2 for gas with temperatures less than $10^4$K is, therefore, exacerbated by the underestimate of metal line cooling.

We estimated the maximum amount of cooling that could theoretically be produced by CII and compared it to the cooling rates of the gas in DH2 (Figure~\ref{fig:coolcurve}).
To make this estimate, we assumed that all carbon in the gas below $10^4$K was in the CII state and that CII cools at a rate of $2.89\times10^{-20}$ N(CII)/N(H) erg s$^{-1}$ H$^{-1}$  \citep{Lehner04}.
We calculated the amount of carbon from the oxygen abundances for each gas particle using the equation for C/O from \citet{Garnett95}.
At the low metallicities of the gas, we found that the estimated maximum amount of cooling from CII is approximately equal to the cooling from $\Hmol$ in regions with molecular gas.

In the future, extending our calculations of the ISRF and shielding to include chemicals other than $\Hmol$ will allow us to better model the cooling of dense gas in the ISM.
At present, though, the introduction of $\Hmol$ in combination with shielding is an important step in the simulation of the cold ISM.
The addition of shielding and efficient low temperature cooling will quickly lead to the creation of cold gas in regions shielded from photoheating, regardless of the source or exact rate of that cooling.
Essentially, so long as heating is reduced by shielding, the presence of low-temperature cooling will lead to the creation of cold gas.
Therefore, to the extent that the cold gas is resolved, this implementation of $\Hmol$ leads to a good first approximation of what the ISM would be like in simulations with more advanced models of low-temperature cooling.

%%%%%%%%%%%%%% Commented Out %%%%%%%%%%%%%%%%%
\begin{figure}
\centering
\includegraphics[width = 0.5\textwidth]{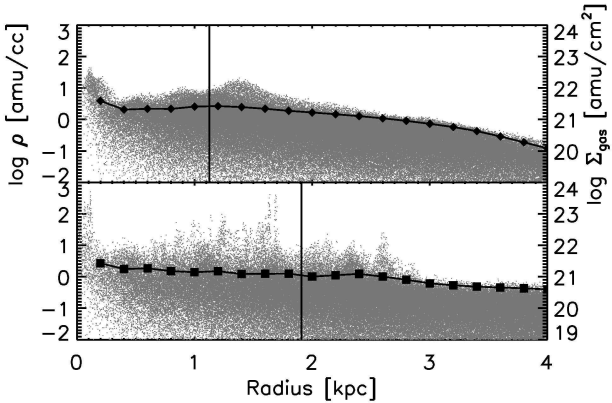}
\caption[Gas density as a function of radius for simulations of dwarf galaxies with and without $\Hmol$]
{Gas density (left vertical axes and points) and radially averaged surface density (right vertical axes and lines) as a function of radius for DnoH2 (top) and DH2 (bottom).  The radially averaged surface density is similar in both galaxies and slightly higher in DnoH2.  The spread in densities at a given radii is higher in DH2 than in DnoH2, indicating a clumpier ISM. }
\label{fig:sd_dens}
\end{figure}

The distribution of dense gas in both DH2 and DnoH2 can give insight in to how changes to the thermodynamics affected the amount and distribution of star formation.
In DnoH2, star formation was mostly confined to the center of the galaxy by a redshift of zero.  
In contrast, when $\Hmol$ was included star formation continued to occur throughout the disk.
As can be seen in Figure~\ref{fig:sd_dens}, this effect does not seem to be due to higher average surface densities at large radii.
The surface density of DnoH2 is, in fact, slightly higher than in DH2 within $r_{25}$.
However, individual gas particles in DH2 reach high densities at larger radii than in DnoH2.
The spikes in Figure~\ref{fig:sd_dens} represent clumps of dense, cold gas in the disk.
This clumpier ISM was the result of the formation of cold gas in DH2.
In that simulation, high-density clumps were able to form throughout the disk, whereas in DnoH2 the only area with high density gas particles was at the center of the galaxy.

The clumpier ISM in DH2 also affected the amount of star formation at z = 0.
Because more of the gas in DH2 was capable of star formation than in DnoH2, star formation continued to occur at a higher rate.
Broad conclusions about the effect of the inclusion of $\Hmol$ on star formation in galaxies of different masses and environments cannot be drawn from one simulation.
However, this test galaxy illustrates how the incorporation of $\Hmol$ results in a clumpier ISM and the level of importance that effect can have to modeling star formation in general. 
In the future, we will increase our set of simulated galaxies in order to study $\Hmol$ within the context of different galactic properties.

In addition to changing the distribution of star formation at z = 0, the ability of gas to reach colder temperatures in DH2 is also potentially responsible for the increased mass in the disk of the galaxy.
Both DnoH2 and DH2 accreted the same fraction of their gas over their lifetime and a slightly higher fraction of gas was expelled from DH2.
This increased gas loss is the effect of a clumpier ISM  \citep{Governato10, Brook10, Brook11a}.
However, a greater mass of gas was stripped form the halo of DnoH2 than in DH2, possibly because the greater ability of gas in DH2 to cool led to the gas being more tightly bound.

By linking star formation of $\Hmol$, a metallicity dependency is introduced into the star formation law.
As metallicity decreases, the lack of shielding causes $\Hmol$ and, consequentially, star formation to take place in higher density gas.
The result is an effective star formation threshold that shifts with metallicity -- something that fixed density thresholds cannot produce \citep{Kuhlen11}.
We observe the HI-to-$\Hmol$ transition to shift to higher density with decreasing metallicity in our model of an isolated MW galaxy.
Similarly, the resolved K-S relation for the simulated dwarf galaxies fits with that of low-metallicity dwarf galaxies.
Finally, we find that the simulated dwarf galaxies, like observed ones, lie below the global K-S relation for disk galaxies, indicating a decreased star formation efficiency on global scales compared to disk galaxies.
This decrease is the result of smaller fractions of star forming gas, either because the lower metallicity causes an effective increase in the star formation threshold or because these galaxies have a lower fraction of high surface density gas.

We found a lower fraction of the baryonic mass in the form of stars in DH2 but the overall increase in the amount of dense, disk gas led to an increase in the global SFR.
While DH2 had a slightly {\em decreased} star formation efficiency, the increased amount of baryons led to a somewhat higher stellar mass.
At high redshift when the metallicities were even lower, DH2 forms only slightly fewer stars than DnoH2.
This result initially appears contradictory to the simulations presented in \citet{Kuhlen11}.
In \citet{Kuhlen11}, equilibrium $\Hmol$ abundances were used to calculate star formation in simulations of dwarf galaxies evolved to z = 4, resulting in dramatically decreased stellar mass.
The simulations presented here differ from those in \citet{Kuhlen11} in the use of a non-equilibrium $\Hmol$ calculation, a lack of a minimum $\Hmol$ fraction for star formation, which allows for star formation in a greater variety of environments, and the shielding of $\Hmol$ and HI gas from photoheating.
We also included a more efficient model for SN feedback, in comparison to \citet{Kuhlen11}.
The latter difference is a particularly important.

In \citet{Kuhlen11} efficient SN feedback was intentionally not included in order to isolate the effects of an $\Hmol$-dependent threshold.
In our simulations, the efficient feedback is included to more closely simulate the behavior of gas in actual galaxies.
Such feedback is responsible for removing baryons from the galaxy, resulting in an appropriate fraction of baryon mass in the disk and an appropriate central concentration of the galaxy.
The result of including efficient feedback is that the feedback is by far the strongest regulator of star formation.
In our simulations, the balance between the heating of gas by feedback and the cooling of gas in the galaxy is more important to setting the global SFR than the change in star formation efficiency produced by an $\Hmol$-based star formation recipe.
We, therefore, most strongly see the effect of $\Hmol$ on star formation indirectly through how it influences the amount of gas in the disk and where the cold, dense star-forming gas is found.
Given the differences in $\Hmol$ model, star formation, thermodynamics, and feedback, it is unsurprising that we do not witness the same dramatic decrease in star formation.

% ----------------- Connection between SF and H2 ------------------------
Basing our star formation recipe on the abundance of $\Hmol$ relies on the premise that the transition to $\Hmol$ is necessary before star formation can begin.
The idea that the formation of $\Hmol$ is a necessary precursor to star formation has been driven in part by the large number of observations of spiral galaxies linking CO emission (and by extension, $\Hmol$) to star formation and has been assumed or argued in a number of theoretical papers \citep{Elmegreen07, RobertsonKravtsov08, Gnedin09, Pelupessy09, Krumholz09a}.
However, others suggest that $\Hmol$ is not necessary for star formation but instead merely correlated with the conditions that lead to star formation in normal spiral galaxies. 
In this model, star formation is initiated independent of whether the hydrogen is atomic or molecular \citep[e.g.][]{Li05,Ostriker10,MacLow10}.
Meanwhile, the same processes that trigger star formation, such as converging gas flow, cloud collisions, and large-scale gravitational collapse, would generally also result in $\Hmol$ in spiral galaxies.

Assuming, though, that star formation occurs because of changes in the gas related to the formation of $\Hmol$, the question remains as to why this transition is important.
 \citet{Schaye04} and \citet{Krumholz11} argue that the fundamental step that leads to star formation is the transition to lower temperatures that happens concurrently with the transition from atomic to molecular hydrogen.
In a similar model, \citet{Glover11a} argued that the formation of gas shielded by dust is what leads both to the formation of $\Hmol$ and to the formation of cold gas that leads to star formation. 
While recognizing the importance of this debate, we have proceeded with our research under the assumption that $\Hmol$ or the shielding and low temperatures that accompany it are intrinsically linked to star formation.
In the future, the ability to alter the dependency of star formation on $\Hmol$ in models of galaxy formation could be important for testing the relationship between the two.

A second area of research that warrants further investigation is the connection between $\Hmol$ and carbon monoxide, CO.
Because of the faintness of $\Hmol$ emission at the temperatures of molecular clouds, CO is frequently used as a tracer for $\Hmol$ in extragalactic observations.
The velocity-integrated CO line intensity ($W_{\mathrm{CO}}$) is converted to a molecular gas column density $N_{\Hmol}$ using the conversion factor, $X_{\mathrm{CO}}$.
Within the Milky Way and similar spiral galaxies, $X_{\mathrm{CO}}$ appears largely consistent.
In lower metallicity environments, such as dwarf and high-z galaxies, or in high-surface density environments, such as the Galactic Center and sub-millimeter galaxies, $X_{\mathrm{CO}}$ is thought to deviate strongly from the standard value.

Because of its importance in observations, $X_{\mathrm{CO}}$ as a function of environment is an active area of observational research \citep[e.g.][[]{Israel97, Leroy09, Gratier10, Leroy11a}.
Simulations of CO and $\Hmol$ have also been used to asses the variation of $X_{\mathrm{CO}}$ with the local gas properties.
In a series of papers including \citet{Glover10}, \citet{Glover11} and \citet{Shetty11a, Shetty11b}, $X_{\mathrm{CO}}$  was studied using magneto-hydrodynamic simulations that followed the chemical network of CO and $\Hmol$ within turbulent molecular clouds.
Simulating CO requires high resolution and detailed radiative transfer but can be done assuming equilibrium \citep{Wolfire10, Glover11}.
Therefore, in galaxy scale simulations it is often calculated during post-processing.
In \citet{Feldmann12}, the CO model from small-scale simulations was combined during post-processing with simulations of galaxies computed with $\Hmol$ as in \citet{Gnedin11}.
\citet{narayanan11} and \citet{Narayanan12} combined detailed radiative transfer and equilibrium models of CO and $\Hmol$ in the post-processing of isolated disk galaxies and mergers.
By including a model for CO abundance in our simulations as done in these works, we would be better able to compare the molecular gas in our simulation to recent observations of molecular gas in galaxies.
Given the complexity of the the problem and the necessity of invoking an additional model for CO during post-processing, we have left the study of CO to future work.

%----------------------------------------------------
\section{Conclusion} \label{conclusionsec4}
In this work, we described a method for calculating the non-equilibrium abundance of $\Hmol$ in SPH cosmological simulations.
Using this method, we were able to reproduce the observed transition from atomic to molecular hydrogen as a function of column density.
We also showed that dependence of $\Hmol$ abundance on volume density and metallicity is consistent with previous studies.
We demonstrated that this method produces consistent results for gas particle masses of approximately $10^4 \Msun$ and less.
The inclusion of $\Hmol$ and shielding enabled gas to reach lower temperatures, which resulted in us being able to simulate the cold phase of the ISM.
In addition to changes to the thermodynamics of the gas, the inclusion of $\Hmol$ allowed us develop a more physically motivated star formation recipe in which star formation is directly linked to the local $\Hmol$ abundance.
By making star formation dependent on $\Hmol$, star formation also became dependent on the local metallicity and LW radiation.

Using this new method, we created the first cosmological simulation with $\Hmol$ of a galaxy integrated to a redshift of zero.
This dwarf galaxy with $\Hmol$ was compared to a galaxy produced without $\Hmol$, in order to study the effect the inclusion of $\Hmol$ has on the  stellar and gaseous content.
We found that when $\Hmol$ was included in the simulation, a larger baryonic mass was in the disk of the galaxy at a redshift of zero, resulting in a brighter and more gas rich galaxy.
The inclusion of $\Hmol$ also resulted in more star formation at later times and a bluer galaxy.
Finally, the simulation with $\Hmol$ had a larger distribution of star formation.

These changes can all be related back to changes to the ISM temperature and density distributions.
The presence of shielded gas and, to a lesser extent, the additional $\Hmol$ cooling resulted in the increased formation of cold gas.
It also resulted in a clumpier ISM and allowed regions dense enough for star formation to form at larger radii.

These simulations demonstrate the importance of modeling $\Hmol$.
Changes to the thermodynamics of the ISM have a strong effect on the mass content and on the spatial and temporal distribution of star formation.
The spatial distribution of stars is additionally important because it can affect both the baryonic and DM distribution of mass through feedback.

In future work, we will examine how the interaction between feedback, star formation, and $\Hmol$ affect disk properties and star formation in different masses of galaxies and at different redshifts.
While they are sensitive laboratories for star formation, dwarf galaxies have only small amounts of $\Hmol$.
Larger, more metal rich galaxies contain more $\Hmol$ and so simulations of them could be more affected by the larger fraction of cold, clumpy gas.
We, therefore, intend to study the effect $\Hmol$ on star formation and feedback over a range of galactic masses. 
Furthermore, a more in-depth examination simulations of galaxies with $\Hmol$ at other redshifts and across their evolution will give insight as to how the evolution of the gas metallicities and surface densities and the accretion of matter affect the presence of $\Hmol$ and $\Hmol$-based star formation.  

%----------------------------------------------
\section{Acknowledgments}
We thank the anonymous referee for the insightful comments.
C.C. was partially supported through the National Science Foundation Graduate Research Fellowship Program and the NSF via grant AST-1009452.
C.C, T.Q. and F.G. were partially supported by NSF AST 0908499.
The Condor Software Program (Condor), on which some of these simulations were run, was developed by the Condor Team at the Computer Sciences Department of the University of Wisconsin-Madison.  All rights, title, and interest in Condor are owned by the Condor Team.  
Larger simulations were run on National TeraGrid machines and NASA AMES and at the Texas Supercomputing Center.

%--------------------------------------------------------------------------
\appendix
\section{Isolated Simulations}
\label{sec:isoIC}

%---------------------------------- Introduction and motivation ----------------
We used an isolated Milky Way-like disk galaxy (virial mass equal to $10^{12}\Msun$) to test our implementation of $\Hmol$ formation and destruction at varying resolutions and metallicities and to test our estimate of the local LW flux.
Isolated simulations have the advantage that they can be simulated at much higher mass resolution than cosmological simulations of similar mass galaxies.  
We choose an approximately Milky Way-mass simulation in order to compare our results to observations of the Milky Way.

In order to test our $\Hmol$ implementation, we created the initial conditions for a set of Milky Way-like disk galaxies of different mass-resolutions and metallicities in the following manner.
Prior to the addition of $\Hmol$, we allowed a stable disk galaxy to form from a DM halo containing a gas cloud, as in \citet{Kaufmann07} and \citet{Stinson06}.
To do this, we began from a live, equilibrium Navarro-Frenk-White (NFW) DM halo \citep{Navarro97} containing a cloud of gas.
The cloud of gas initially shared the same NFW profile as the DM, had a uniform rotational velocity, and had a temperature distribution such that it was in hydrostatic equilibrium prior to cooling.
We then integrated the simulation for $1.0$ Gyr without $\Hmol$ while allowing the gas cloud to collapse through ionization and metal-line cooling and to form stars according to our non-$\Hmol$ star formation recipe.
After 1.0 Gyr, a stable galaxy with a stellar and gaseous disk had formed inside the DM halo.  
We adjusted the metallicity of the gas to reflect the range of metallicities we were interested in examining by adding a constant metallicity offset to the disk gas.
We chose the metallicity offset such that the mean metallicity of the gas in the disk of the galaxy equaled in turn 1.0, 0.3, and 0.1 Z$_{\sun}$.

\begin{table*}
\begin{minipage}{100mm}
\caption{Isolated Milky-Way like simulations used to test the code and their parameters.
When calculating the mean smoothing length, $\langle h \rangle$, of the particles in the disk, we selected particles with densities of 0.1 amu/cc or greater.
The gravitational softening length was $206 \pc$ in all of the simulations.}
\label{tab:isoSims_H2}
\begin{center}
\begin{tabular}{lccccr}
\hline
& \multicolumn{3}{c}{Particle Mass [$\Msun$]} &   & \\
\cline{2-4}\\
Name & Dark & Star & Gas & $\langle h \rangle$ & Mean Metallicity \\
 &  &  & & in Disk [pc] & in Disk [$Z/Z_{\sun}$] \\ \hline
MWlr                & $10^7$    &  $4.0 \times 10^5$     & $1.4 \times 10^6$           & 278       & 1 \\ 
MWmr              & $10^6$    &  $4.0 \times 10^4$     & $1.4 \times 10^5$           & 176       & 1\\
MWmr0.3z       & ``              & ``                                    & ``                                    & ``          & 0.3\\  
MWmr0.1z       & ``              & ``                                    & ``                                    & ``          & 0.1\\  
MWhr               & $10^5$    &  $4.0 \times 10^3$      & $1.4 \times 10^4$           &  82          & 1\\  
\end{tabular}
\end{center}
\end{minipage}
\end{table*}

The set of disk galaxies of varying metallicities and mass resolutions created using the above methodology became the initial condition for our simulations used to test the $\Hmol$ implementation.
Table~\ref{tab:isoSims_H2} contains a list of these galaxies and their parameters.
For each of these galaxies, we enabled $\Hmol$ in the galaxy and continued the simulations for a further 100 Myrs. 
This time period was chosen to be sufficiently larger than the star and $\Hmol$-formation time scales that the gas and star formation in the disk could reach equilibrium.

%----------------------------------------------
\bibliographystyle{apj} 
\bibliography{./MolecH_Methods_paper}

\end{document}